\documentclass[pra,twocolumn,superscriptaddress,showpacs,amsmath,amstex,amssymb,citeautoscript]{revtex4-1}

\usepackage[T1]{fontenc}
\usepackage{natbib}
\usepackage[utf8]{inputenc}
\usepackage{lipsum}
\usepackage{amsmath}
\usepackage{amssymb}
\usepackage{bbm}
\usepackage{braket}
\usepackage{xcolor}
\usepackage{pifont}
\usepackage[mathscr]{euscript}
\usepackage[shortlabels]{enumitem}
\usepackage[justification=justified]{subcaption}
\usepackage{graphicx}
\usepackage{lipsum}
\DeclareMathOperator{\tr}{Tr}
\allowdisplaybreaks
\usepackage{float}
\usepackage{graphicx}
\usepackage{dsfont}
\usepackage{comment}
\usepackage[colorlinks=true]{hyperref}  
\hypersetup{
    bookmarks=true,         
    unicode=false,          
    pdftoolbar=true,        
    pdfmenubar=true,        
    pdffitwindow=false,     
    pdfstartview={FitH},    
    pdftitle={Physics-informed Transformers},    
    pdfauthor={Sobral, Perle, Scheurer},     
    pdfsubject={},   
    pdfcreator={},   
    pdfproducer={}, 
    pdfkeywords={} {} {}, 
    pdfnewwindow=true,      
    colorlinks=true,       
    linkcolor=blue, 
    citecolor=blue,        
    filecolor=magenta,      
    urlcolor=blue           
}

\newcommand{\equref}[1]{Eq.~(\ref{#1})}

\newcommand{\secref}[1]{Sec.~\ref{#1}}
\newcommand{\figref}[1]{Fig.~\ref{#1}}

\newcommand{\appref}[1]{Appendix~\ref{#1}}

\newcommand{\pdagger}{{\phantom{\dagger}}}

\newcommand{\sign}{\,\text{sign}}
\renewcommand{\approx}{\simeq}
\renewcommand{\Re}{\text{Re}}

\renewcommand{\vec}[1]{\boldsymbol{#1}}

\definecolor{wrongultramarine}{rgb}{1,0.5,0}

\begin{document}

\title{Physics-informed Transformers for Electronic Quantum States}

\author{João Augusto Sobral}
\affiliation{Institute for Theoretical Physics III, University of Stuttgart, 70550 Stuttgart, Germany}
\author{Michael Perle}
\affiliation{Institute for Theoretical Physics, University of Innsbruck, A-6020, Innsbruck, Austria}
\author{Mathias S.~Scheurer}
\affiliation{Institute for Theoretical Physics III, University of Stuttgart, 70550 Stuttgart, Germany}

\begin{abstract}
Neural-network-based variational quantum states in general, and more recently autoregressive models in particular, have proven to be powerful tools to describe complex many-body wave functions. However, their performance crucially depends on the computational basis chosen and they often lack physical interpretability. To mitigate these issues, we here propose a modified variational Monte-Carlo framework which leverages prior physical information to construct a computational second-quantized basis containing a reference state that serves as a rough approximation to the true ground state. In this basis, a Transformer is used to parametrize and autoregressively sample the corrections to the reference state, giving rise to a more interpretable and computationally efficient representation of the ground state. We demonstrate this approach using a non-sparse fermionic model featuring a metal-insulator transition and employing Hartree-Fock and a strong-coupling limit to define physics-informed bases. We also show that the Transformer’s hidden representation captures the natural energetic order of the different basis states. This work paves the way for more efficient and interpretable neural quantum-state representations.
\end{abstract}

\maketitle

\section{Introduction}
Neural quantum states (NQS) have been successfully used within Variational Monte Carlo (VMC) to describe highly accurate and flexible parametrizations of the ground state wavefunction of a variety of many-body physical systems \cite{carleoSolvingQuantumManybody2017b, Pfau2020Sep, Valenti2022Jan, hermannInitioQuantumChemistry2023, medvidovicNeuralnetworkQuantumStates2024c, langeArchitecturesApplicationsReview2024b, Melko2024Jan}. Parallel developments have expanded NQS capabilities to capture excited states \cite{Choo2018Oct, pfauAccurateComputationQuantum2024a}, while improvements of the stochastic reconfiguration method \cite{chenEmpoweringDeepNeural2024a, Rende2024Aug} have enhanced both the scalability and accuracy of these variational ansätze. Recently, hybrid approaches which integrate NQS with experimental or computational projective measurements in a pre-training stage \cite{Czischek2022May, Moss2024Mar,langeTransformerNeuralNetworks2024, Ibarra-Garcia-Padilla2024Nov}, or quantum-classical ansätze \cite{Barison2023Sep, Metz2024Sep} have also shown substantial VMC performance improvements.

Neural autoregressive quantum states (NAQS), which are based on the idea of efficiently parameterizing joint distributions as a product of conditional probabilities, have acquired substantial attention due to their general expressiveness and ability to perform efficient and exact sampling \cite{medvidovicNeuralnetworkQuantumStates2024c, sharir2020autoregressive}. Recurrent Neural Networks \cite{Elman1990Mar, Lipton2015May}  and Transformers \cite{Vaswani2017Jun, Lin2021Jun} constitute prominent examples of autoregressive architectures commonly used as variational ansätze \cite{Hibat-Allah2020Jun, Lange2024Jun, Luo2019Jun, zhang2023tqs, viterittiTransformerVariationalWave2023a}. 
Transformer quantum states (TQS), in particular, have proven effective in providing highly accurate representations of ground states in frustrated magnetism \cite{viterittiTransformerVariationalWave2023a, viterittiTransformerWaveFunction2024} 
and Rydberg atoms \cite{spragueVariationalMonteCarlo2024}, while also holding promise for interpretability within the context of the self-attention mechanism \cite{cheferTransformerInterpretabilityAttention2021,Cui2024Feb, rendeMappingAttentionMechanisms2024, Rende2024Oct}. 

Despite their versatility, NQS effectiveness may still depend on the basis in which the Hamiltonian is represented. For instance, Robledo-Moreno \textit{et al.} \cite{morenoEnhancingExpressivityVariational2023b} demonstrated that variationally optimized single-particle orbital rotations can significantly improve the accuracy of calculated observables. 
Furthermore, NQS wave function representations may lack direct physical interpretability, e.g., with respect to the relative frequency of sampled states from the Hilbert space. This contrasts with post-Hartree-Fock (HF) methods in quantum chemistry, such as coupled cluster theory \cite{Bartlett2007Feb}, where corrections are naturally interpreted as single or double excitations to the HF state.

We present a modified VMC approach that simultaneously addresses these aspects. Although the method is architecture-agnostic, we demonstrate its effectiveness using a TQS \cite{zhang2023tqs, viterittiTransformerVariationalWave2023a}, given their previously mentioned advantages. As a first step, an energetically motivated basis is chosen (see also \figref{fig:generalmethodology}a); for concreteness, we here use two examples---the basis that diagonalizes the Hamiltonian in the mean-field approximation and a natural basis in the limit of strong interactions of our model. Both of these bases contain a ``reference state’’ (RS) which is a candidate for an approximate description of the ground state of the system. In the case of the mean-field approximation, the RS just corresponds to the Hartree-Fock (HF) ground state. Meanwhile, for the second basis, the RS is the exact ground state at strong coupling. We explicitly parametrize the weight of the RS using a single parameter $\alpha \in \mathbb{R}$ while the Transformer network focuses on describing the corrections to it. Apart from enhancing convergence, $\alpha$ is convenient as it directly quantifies how close the many-body state is to the interpretable RS. We emphasize that this approach (as opposed to, e.g., coupled cluster methods) is not biased toward favoring states close to the RS or, equivalently, $\alpha$ near $1$. In fact, we demonstrate explicitly that the technique leads to vanishingly small weight of the RS should the latter not be a good approximation to the true ground state. What is more, for instance in the HF basis, also the remaining basis states have a natural interpretation as being associated with a certain number of particle-hole excitations in the HF bands, yielding a natural energetic hierarchy that we also recover in their relative weight and hidden representation of the Transformer’s parameterization of the many-body ground state. 

To exemplify this methodology, we use a one-dimensional interacting fermionic many-body model in momentum space. This model features an exactly solvable strong-coupling limit,  which is used to define the strong-coupling basis mentioned above. Moreover, it exhibits a finite regime where integrability is no longer apparent, showing clear differences between exact diagonalization (ED) and HF, where corrections to mean-field treatments become significant.

Our results demonstrate that when the true ground state is close to a product state (the strong coupling limit), the HF basis (strong coupling basis) guides the TQS to converge to a variational representation with two key characteristics: (i) 
the number of states required for an accurate ground state representation only involves a fraction of the total Hilbert which is learned and efficiently sampled from by the transformer; (ii) the states self-organize hierarchically by their statistical weights, with a clear physical structure on latent space, naturally representing excitations on top of the RS. Finally, we show how these features contrast sharply with a generic basis, which generically requires an exponentially large amount of states, hindering scalability and the identification of dominant corrections to mean-field treatments. 

\section{Results}
\label{sec:results}
\subsection{General Formalism}\label{GeneralFormalism}
Our central goal is to determine the ground state of a general interacting fermionic Hamiltonian $\hat{H}$ given by

\begin{figure}[t!]
    \centering
    \includegraphics[width=0.49\textwidth]{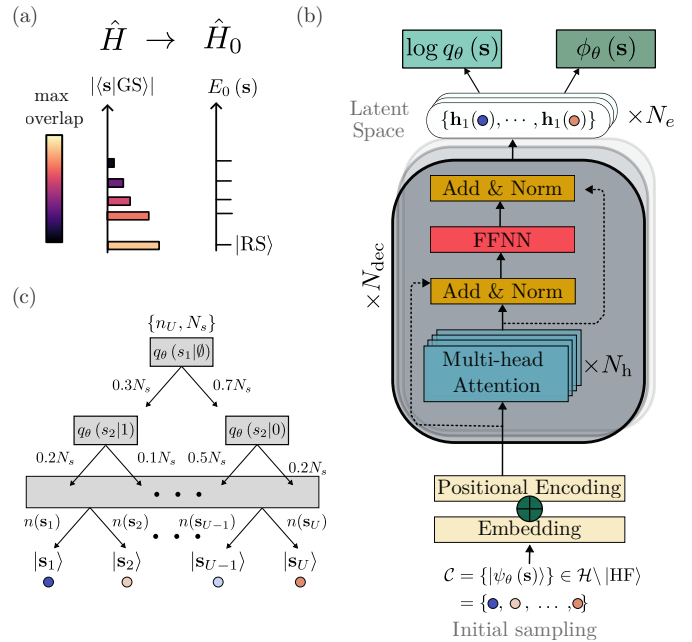}
    \caption{\textbf{General methodology.} (a) First, we choose a $\hat{H}_{0}$ approximating the target Hamiltonian $\hat{H}$, e.g., via a mean-field approximation or by taking the strong-coupling limit. We use the groundstate $\ket{\text{RS}}$ and excited states $\ket{\vec{s}}$ of $\hat{H}_{0}$ to define a physics-informed, interpretable basis for the transformer (b) in \equref{eq:tqsmod}; as long as the dominant weight of the ground state of $\hat{H}$ is in the low-energy part of the spectrum $E_0(\vec{s})$ of $\hat{H}_{0}$, this further improves sampling efficiency and the expressivity of the ansatz.       
    (c) We sample the states $\vec{s}$ using the batch-autoregressive sampler \cite{barrettAutoregressiveNeuralnetworkWavefunctions2022b,malyshev2023autoregressive, Malyshev2024Aug}.
    It is controlled by the batch size $N_{s}$ and the number of partial unique strings $n_{U}$, and directly produces the relative frequencies $r(\vec{s})$ associated with each state in a tree structure format. Back to (b), the states $\vec{s}$ are then mapped to a high-dimensional representation of size $d_{
    	\text{emb}}$ and passed through $N_{\text{dec}}$ decoder-layers \cite{zhang2023tqs}, containing $N_{h}$ attention heads, which produce correspondent representations $\vec{h}\left(\vec{s}\right)\in \mathbb{R}^{d_{\text{emb}}}$ in latent space. As discussed in the main text, the wavefunctions $\psi_{\vec{\theta}}(\vec{s})=\sqrt{q_{\vec{\theta}}(\vec{s})}e^{i\phi_{\vec{\theta}}(\vec{s})}$  can be directly obtained from these vectors. A new set of states $\mathcal{C}$  is then obtained, according to the updated $q_{\vec{\theta}}(\vec{s})$, and the process is repeated until the convergence of $\{\vec{\theta},\alpha\}$ according to \equref{eq:loss}.}
    \label{fig:generalmethodology}
\end{figure}
\begin{align}\begin{split}
\hat{H}=&\sum_{a, b ,\vec{k}} d_{\vec{k}, a}^{\dagger} h_{a, b}(\vec{k})d_{\vec{k},b}^{\pdagger} +\\& +
\sum_{\substack{a_1,a_2,b_1,b_2 \\ \vec{k}_{1},\vec{k}_{2},\vec{k}_{3},\vec{k}_{4}}} d_{\vec{k}_{1},a_1}^{\dagger} d_{\vec{k}_{2},a_2}^{\dagger} d_{\vec{k}_{3},b_2}^{\pdagger} d_{\vec{k}_{4},b_1}^{\pdagger} V_{a_1,a_2,b_2,b_1}^{\vec{k}_{1},\vec{k}_{2},\vec{k}_{3},\vec{k}_{4}},
\label{eq:generalhamiltonian}
\end{split}\end{align}
where $d^\dagger_{\vec{k},a}$ and $d_{\vec{k},a}$ are fermionic, second quantized creation and annihilation operators with momentum $\vec{k}$, and indices $a,b,\dots$ indicate additional internal degrees of freedom of the system, such as spin and/or bands. The one and two-body terms are determined by $h_{a,b}(\vec{k})$ and $ V_{a_1,a_2,b_2,b_1}^{\vec{k}_{1},\vec{k}_{2},\vec{k}_{3},\vec{k}_{4}}$, respectively; although not a prerequisite for our method, we assume translational invariance for notational simplicity. 

A first approximation to the ground state of \equref{eq:generalhamiltonian} can be provided by HF \cite{Carsky1991Jul, Fukutome1981Nov}; restricting ourselves  to translation-invariant Slater-determinants, HF can be stated as finding the momentum-dependent unitary transformations $U_{\vec{k}}$ of the second-quantized operators,
\begin{equation} 
\bar{d}_{\vec{k}, p}=\sum_{a}\left(U_{\vec{k}}\right)_{p,a} d_{\vec{k}, a},
\label{eq:basistransfo}
\end{equation}
such that the HF self-consistency equations are obeyed  (see \appref{HFAppendix}) and the Hamiltonian assumes a diagonal quadratic form within the mean-field approximation, i.e.,
\begin{equation}
\hat{H}=\sum_{\vec{k}, p} \epsilon_{\vec{k}, p} \bar{d}_{\vec{k}, p}^{\dagger} \bar{d}^\pdagger_{\vec{k}, p}+\ldots,
\label{eq:unitarytransf}
\end{equation}
where the ellipsis indicates terms beyond mean-field. The transformations in \equref{eq:basistransfo} are obtained in an iterative approach until a specified tolerance is reached. 

The ground state within HF is given by filling the lowest fermionic states in \equref{eq:unitarytransf}, which we will use as our RS, denoted by $\ket{\text{RS}}$ in the following. Importantly, though, HF also defines an entire basis via \equref{eq:basistransfo}, which is approximately related to the Hamiltonians spectrum as parametrized by $\epsilon_{\vec{k},p}$.
We leverage both the spectrum $\epsilon_{\vec{k},p}$ and its associated basis to improve sampling efficiency and physical interpretability within the VMC framework. As summarized graphically in \figref{fig:generalmethodology}(a,b), we express the many-body state in the HF basis (\ref{eq:basistransfo}) and denote the associated computational basis by $\ket{\vec{s}}$, where $\vec{s}=\left(s_{1},\dots,s_{N_{k}}\right)$ labels the occupations of the fermionic modes created by $\bar{d}_{\vec{k}, p}^{\,\dagger}$ in the $N_{\vec{k}}$ different electronic momenta $\vec{k}$. Our variational many-body ansatz then reads as
\begin{equation}
    \left|\Psi_{\{\vec{\theta},\alpha\}}\right\rangle=
  \alpha\left|\text{RS}\right\rangle+\sqrt{1-\alpha^2}
     \sum_{\vec{s} \neq \text{RS}} 
     \psi_{\vec{\theta}}(\vec{s})|\vec{s}\rangle,
     \label{eq:tqsmod}
\end{equation}
where $\psi_{\vec{\theta}}(\vec{s}) \in \mathbb{C}$ is a neural network representation \cite{carleoSolvingQuantumManybody2017b} of the amplitudes for the states $\vec{s}$ that are not the RS, and $\alpha$ is an additional variational parameter describing the weight associated to the RS. Note that a global phase choice allows us to take $\alpha \in \mathbb{R}$ without loss of generality.

The motivation for the variational parameter $\alpha$ is two-fold. First, it explicitly quantifies deviations of the ground state from the RS, which for HF refers to the optimal product state. A ground state being close to the RS is then reflected by $\alpha$ approaching unity, while small $\alpha$ will indicate strong deviations from a product state. As such, our approach combines the interpretability of HF with the lack of being constrained to (the vicinity of) a Slater determinant. We emphasize that different HF calculations, e.g., restricted to be in certain symmetry channels, can be used and compared.
Secondly, through \equref{eq:tqsmod}, the NQS can solely focus on the corrections $\delta E$ to the RS energy $E_{\text{RS}}$. 
Since the RS is never sampled by the NQS by construction, this separation is beneficial when HF captures the dominant ground state contributions, as targeting corrections would be hindered by low acceptance probabilities in Metropolis-Hastings sampling \cite{Choo2020May}---a phenomenon analogous to mode collapse in generative adversarial networks \cite{Metz2016Nov, Kanaujia2024May}. If HF is not a good approximation, there is in general no reason why splitting up the contribution of the RS would be detrimental to the network's performance. 

It remains to discuss how the other states, $\vec{s}\neq\text{RS}$, are described through $\psi_{\vec{\theta}}(\vec{s})$ which depends on a set of parameters $\vec{\theta} \in\mathbb{R}^n$. These parameters are jointly optimized with $\alpha$ according to
\begin{equation}
\arg \min_{\{\vec{\theta}, \alpha\}} E(\vec{\theta}, \alpha)=\arg \min_{\{\vec{\theta}, \alpha\}} \frac{\left\langle\Psi_{\{\vec{\theta},\alpha\}}\left|\hat{H}\right| \Psi_{\{\vec{\theta},\alpha\}}\right\rangle}{\left\langle\Psi_{\{\vec{\theta},\alpha\}}\mid \Psi_{\{\vec{\theta},\alpha\}}\right\rangle},
    \label{eq:loss}
\end{equation}
i.e., via a minimization of the energy functional $E(\vec{\theta}, \alpha)$. 
We emphasize that this approach is distinct from neural network backflow \cite{Luo2019Jun} as we use the HF basis to express the many-body state rather than dressing its single-particle orbitals with many-body correlations.
While other approaches are feasible, too, we here employ a Transformer \cite{vaswaniAttentionAllYou2017, zhang2023tqs} to represent the Born distribution $q_{\vec{\theta}}(\mathbf{s})=\left|\psi_{\vec{\theta}}(\vec{s})\right|^{2}/\sum_{\vec{s}^{\prime}}\left|\psi_{\vec{\theta}}(\vec{s}^{\prime})\right|^{2}$ autoregressively, i.e., 
\begin{equation}
q_{\vec{\theta}}(\vec{s})=\prod_{i=1}^{N_{\vec{k}}} q\left(s_i \mid s_{i-1}, \ldots, s_1\right).
\label{eq:autoregressive}
\end{equation}
The associated wave functions, 
$\psi_{\vec{\theta}}(\vec{s})=\sqrt{q_{\vec{\theta}}(\vec{s})}e^{i\phi_{\vec{\theta}}(\vec{s})}
$, can be obtained from the Transformer's latent space (\figref{fig:generalmethodology}b), by taking an affine linear  transformation followed by a softmax activation function for the amplitude, and a scaled softsign activation such that  $\phi_{\vec{\theta}}(\vec{s})\in[-\pi, \pi]$ \cite{zhang2023tqs,Hibat-Allah2020Jun}. The first guarantees that the output of the Transformer gives normalized conditional probabilities in \equref{eq:autoregressive} \cite{sharir2020autoregressive}.

The energy expectation values for the corrections $\delta E$ are calculated as a weighted average over a set $\mathcal{S}$ of $n_U$ unique states (from a batch of $N_{s}$ sampled states $\vec{s}$) from $q_{\vec{\theta}}(\vec{s})$ through the batch auto-regressive sampler \cite{malyshev2023autoregressive, Malyshev2024Aug, Wu2023Jun} (see \figref{fig:generalmethodology}c)  as
\begin{equation}
\langle \delta E\rangle=\mathbb{E}_{\vec{s}\sim q_{\vec{\theta}}}\hspace{-0.3em}\left[{H}_{\mathrm{loc}}(\mathbf{s})\right] \approx \sum_{\mathbf{s} \in \mathcal{S}\neq \text{RS}} {H}_{\mathrm{loc}}(\mathbf{s})r(\mathbf{s}).
\label{eq:averages}
\end{equation}
Here, $r(\mathbf{s})=n(\mathbf{s})/N_{s}$ represents the relative frequency of each state $\vec{s}$ and 
\begin{equation}
{H}_{\mathrm{loc}}(\mathbf{s})=\sum_{\mathbf{s}^{\prime}\neq \text{RS}} \frac{\braket{\mathbf{s}|\hat{H}|\mathbf{s}^{\prime}} \psi_{\vec{\theta}}(\mathbf{s}^{\prime})}{\psi_{\vec{\theta}}(\mathbf{s})}
\label{eq:hloc}
\end{equation}
are the typical local estimators.
According to \equref{eq:tqsmod}, the energy functional is divided into sectors
\begin{align}
E(\vec{\theta}, \alpha)  = 
\alpha^{2}E_{\text{RS}}+ \left(1-\alpha^{2}\right)\mathbb{E}_{\vec{s}\sim q_{\vec{\theta}}}\hspace{-0.3em}\left[{H}_{\mathrm{loc}}(\mathbf{s})\right] +  \nonumber \\ + 
2\alpha\sqrt{1-\alpha^{2}}\Re\left(\mathbb{E}_{\vec{s}\sim q_{\vec{\theta}}}\hspace{-0.3em}\left[{H}^{\text{RS}}_{\text{loc}}\left(\vec{s}\right)\right]\right),
      \label{eq:expenergy} 
\end{align}
with the modified local estimator ${H}^{\text{RS}}_{\text{loc}}\left(\vec{s}\right)=\braket{\vec{s}|\hat{H}|\text{RS}}/\psi_{\vec{\theta}}(\vec{s}) $. 
The network parameters $\vec{\theta}$ are optimized as usual with the gradients of the expression \eqref{eq:expenergy} given by
\begin{equation}
\begin{aligned}
\nabla_{\vec{\theta}} E(\vec{\theta}, \alpha) = 2\Re\left(
 \mathbb{E}_{\mathbf{s}\sim q_{\vec{\theta}}}\hspace{-0.3em}\left[{\mathscr{H}}_{\text{loc}}\left(\vec{s},\alpha\right)\cdot \nabla_{\vec{\theta}}\log \psi^{*}_{\vec{\theta}}\left(\vec{s}\right)\right]\right),  
 \label{eq:gradtheta} 
\end{aligned}
\end{equation}
with 
\begin{equation*}
{\mathscr{H}}_{loc}\left(\vec{s},\alpha\right) = \left( \left(1-\alpha^{2}\right){H}_{\text{loc}}(\vec{s})+
\alpha\sqrt{1-\alpha^{2}}{H}^{\text{RS}}_{\text{loc}}(\vec{s})\right). 
\end{equation*}
To prevent numerical instabilities during the optimization of \equref{eq:tqsmod}, it is necessary to constrain $\alpha$ with the parametrization $\alpha=(1+\tanh{\alpha_{0}})/2$ to the interval $ [-1, 1]$. After updating the network parameters $\vec{\theta}$ at each iteration, the reweighting parameters are dynamically adjusted according to the gradient of $E(\vec{\theta}, \alpha)$ in \equref{eq:expenergy} with respect to $\alpha_{0}$, i.e.,
\begin{equation}
\begin{aligned}
\nabla_{\alpha_{0}}E(\vec{\theta},\alpha) =2\alpha\nabla_{\alpha_{0}}\alpha\left[E_{\text{RS}}-E_{\vec{s}\vec{s}^{\prime}}+\frac{E_{\vec{s}\text{RS}}\left(1-2\alpha^{2}\right)}{2\alpha\sqrt{1-\alpha^{2}}}\right],
\end{aligned}
\label{eq:gradalpha}
\end{equation}
where $E_{\vec{s}\vec{s}^{\prime}}=\mathbb{E}_{\vec{s}\sim q_{\vec{\theta}}}\hspace{-0.3em}\left[{H}_{\mathrm{loc}}(\mathbf{s})\right] $ and $E_{\vec{s}\text{RS}}=2\Re\left(\mathbb{E}_{\vec{s}\sim q_{\vec{\theta}}}\hspace{-0.3em}\left[{H}^{\text{RS}}_{\text{loc}}\left(\vec{s}\right)\right]\right)$. For the optimizer, we use stochastic gradient descent for \equref{eq:gradalpha} and preconditioned gradient methods \cite{Gupta2018Feb,Vyas2024Sep} for \equref{eq:gradtheta} with adaptable learning rate schedulers (see \appref{sec:vmc}).

\subsection{Model Hamiltonian}\label{ModelHamiltonian}
To test and explicitly demonstrate our methodology, we consider a concrete model of the form of \equref{eq:generalhamiltonian}. It describes spinless, one-dimensional electrons which can occupy two different bands, $a = \pm$, as described by the creation and annihilation operators $d^\dagger_{\vec{k},a}$ and $d_{\vec{k},a}$, respectively. They interact through a repulsive Coulomb potential $V(\vec{q})=(2N_{k}\left(1+\vec{q}^{2}\right))^{-1}$. More explicitly, the Hamiltonian reads as 
\begin{equation}
    \hat{H} = t\sum_{\vec{k}\in\text{BZ}} \cos(k) \, d^\dagger_{\vec{k}} \sigma^\pdagger_{z} d^\pdagger_{\vec{k}} + U \sum_{\vec{q} \in \mathbbm{R}} V(\vec{q}) \rho_{\vec{q}} \rho_{-\vec{q}}, \label{eq:hamiltonian} 
\end{equation}
where the momenta $\vec{k}$ are defined on the first Brillouin zone $ (\text{BZ}):= [-\pi, \dots,\pi-2\pi/N_{\vec{k}}]$ of a finite system with $N_{\vec{k}}$ sites and $\sigma_j$ ($j=0,x,y,z$) are the Pauli matrices in band space. The density operator is given by 
\begin{equation}
    \rho_{\vec{q}} = \sum_{\vec{k}\in\text{BZ}} \left(d^\dagger_{\vec{k}+\vec{q}} \mathcal{F}(\vec{k},\vec{q}) d^\pdagger_{\vec{k}} - \sum_{\vec{G}\in\text{RL}} \delta_{\vec{q},\vec{G}} f_{1}(\vec{k},\vec{G})\right),
\end{equation}
where $\text{RL} =2\pi \mathbbm{Z}$ is the reciprocal lattice  and the ``form factors'' read as $\mathcal{F}(\vec{k},\vec{q})=f_1(\vec{k},\vec{q}) + i\sigma_y f_{2}(\vec{k},\vec{q})$; for concreteness, we choose $f_{1}\left(\vec{k},\vec{q}\right)=1$ and  $f_{2}\left(\vec{k},\vec{q}\right)=0.9 \sin \left(\vec{k}\right) \left(\sin \left(\vec{q}\right) +\sin \left(\vec{k}+\vec{q}\right)\right)$ in our computations below. 

Note that this model is non-sparse since all momenta are coupled and, as such, is generally expected to be challenging to solve. It is inspired by models of correlated moir\'e superlattices, most notably of graphene, which exhibit multiple low-energy bands that are topologically obstructed \cite{PhysRevX.8.031089,TBGII}; they can, hence, not be written  as symmetric local theories in real space and are, thus, typically studied in momentum space \cite{PhysRevX.10.031034,PhysRevB.103.205414,Christos2022Apr}. Furthermore, the strong coupling limit, $t/U \rightarrow 0$, of \equref{eq:hamiltonian} can be readily solved: to this end, we introduce a new basis defined by $U_{\vec{k}} = \begin{pmatrix} 1 & -i \\ 1 & i \end{pmatrix}/\sqrt{2}$ in \equref{eq:basistransfo} which diagonalizes the form factors $\mathcal{F}(\vec{k},\vec{q})$ at all momenta. It is then easy to see (see \appref{sec:fm}) that, at half-filling (the number of electrons $N_e = N_{\vec{k}}$), any of $\ket{\pm} = \prod_{\vec{k}} \bar{d}^{\,\dagger}_{\vec{k}, \pm} \ket{0}$ are exact ground states in the limit $t/U \rightarrow 0$; which of the two ground states is picked is determined by spontaneous symmetry breaking: the Hamiltonian is invariant under the anti-unitary operator $PT$ with action $PT\bar{d}_{\vec{k}, \pm}(PT)^\dagger = \bar{d}_{\vec{k}, \mp}$, which is broken by both of these states. The resulting symmetry-broken phase can be shown to exhibit a finite gap.
Importantly, this strong coupling limit defines another natural computational basis and associated $\ket{\text{RS}} = \ket{+}$ or $\ket{-}$, which we will use and compare with the HF basis defined in \secref{GeneralFormalism}; in analogy to twisted bilayer graphene \cite{PhysRevB.103.205414}, we will refer to this strong-coupling basis as ``chiral basis''.

In contrast, at large $t/U$, the non-interacting term in \equref{eq:hamiltonian} dominates and we obtain a symmetry-unbroken metallic phase. As such, there is an interaction-driven metal-insulator transition at half-filling at some intermediate value of $t/U$ ($\approx 0.14$ according to HF). To be able to compare both chiral and HF bases and since half-filling has the largest Hilbert space, we will focus on $N_e=N_{\vec{k}}$ in the following. Furthermore, we will neglect double-occupancy of each of the $N_e$ momenta for simplicity such that the basis states $\ket{\vec{s}}$, with $s_k \in \{0,1\}$, in \equref{eq:tqsmod} can be compactly written as $\ket{\vec{s}} = \prod_{k=1}^{N_{\vec{k}}} \bar{d}_{\vec{k}, (-1)^{s_k}}^{\,\dagger} \ket{0}$.

\begin{figure}[b]
    \centering
    \includegraphics[width=0.49\textwidth]{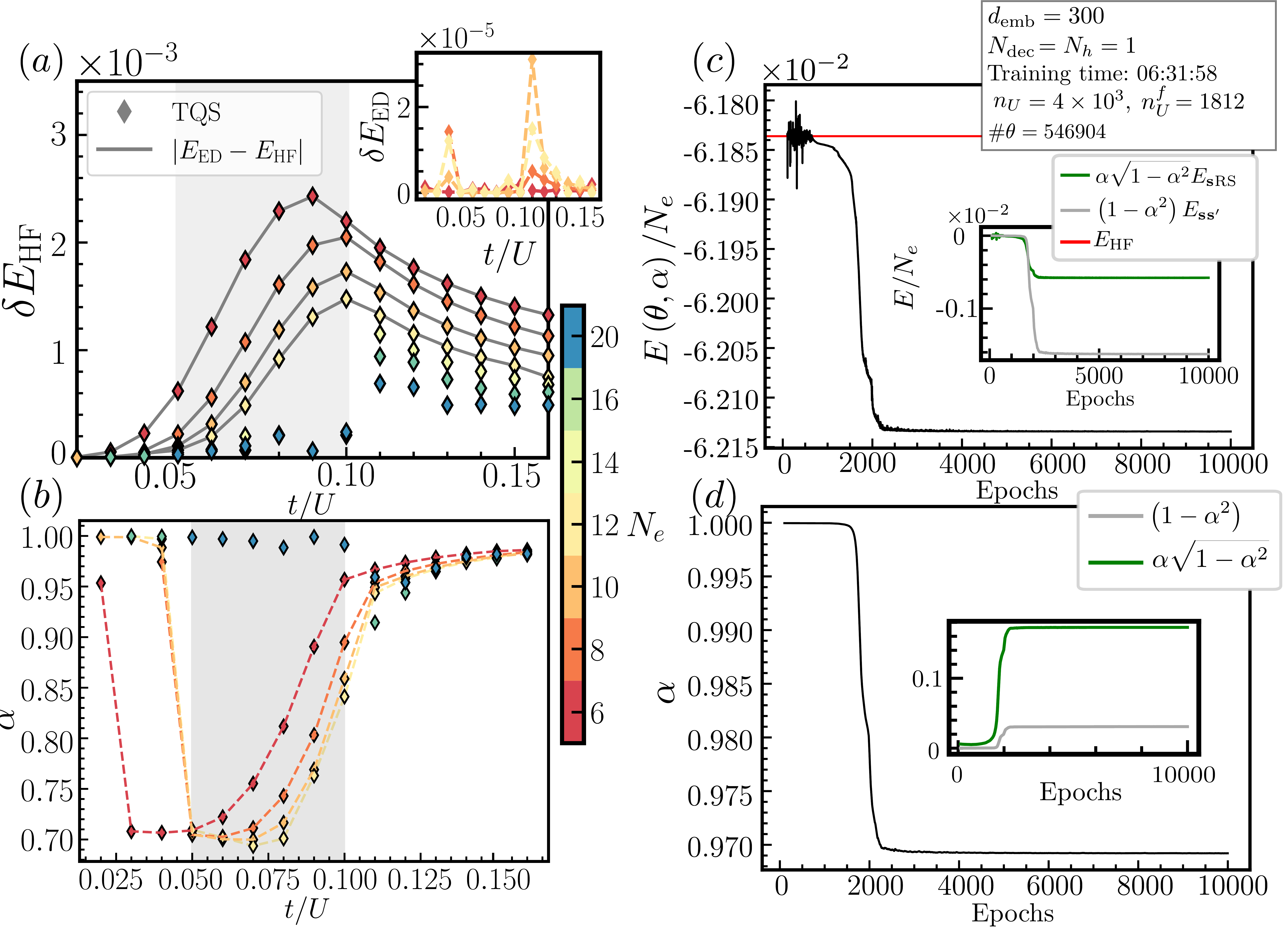}
    \caption{\textbf{Performance of HF-TQS for different system sizes and $t/U$.} (a) Difference between the ground state energy per site and HF for $x$ as ED (solid lines) and TQS (markers) in $\delta E_{\text{HF}}=\left|E_{x}-E_{\text{HF}}\right|$ at various system sizes $N_e$. The inset shows the absolute value of the relative error $\delta E_{\text{ED}}=\left|E_{\text{TQS}}-E_{\text{ED}}\right|$. The corresponding converged $\alpha$ weights [according to \equref{eq:expenergy}] are shown in (b). The gray regions indicate the vicinity of the metal-insulator transition. Convergence of the ground state energy per site (c) and of $\alpha$ (d) as a function of epochs for  $t/U=0.16$ and $N_{e}=30$. $n_{U}^{f}$ is the total number of unique states kept by the Transformer from the value $n_{U}$ set initially in \figref{fig:generalmethodology}(c). Training time refers to one NVIDIA H100 GPU with networks parameters defined in  \figref{fig:generalmethodology}b (see \appref{sec:vmc}). The total number of network parameters $
    \vec{\theta}$ used by the TQS is indicated by $\# \vec{\theta}$.}
    \label{fig:scaling}
\end{figure}

\begin{figure*}[t]
    \centering
     \includegraphics[width=\textwidth]{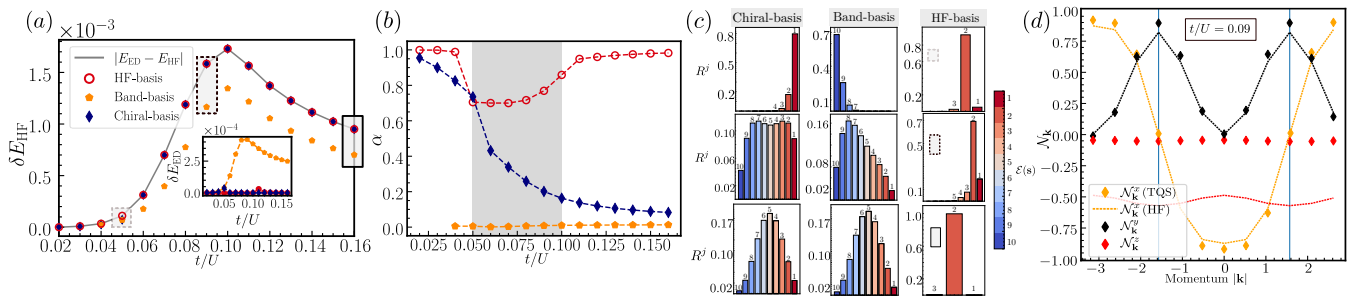}
    \caption{\textbf{Performance comparison of the TQS in different bases.} (a) Difference between the ground state energy per site and HF for $x$ as ED (solid lines) and TQS (markers) in $\delta E_{\text{HF}}=\left|E_{x}-E_{\text{HF}}\right|$ for $N_{e}=10$. The inset shows the absolute value of the relative error $\delta E_{\text{ED}}=\left|E_{\text{TQS}}-E_{\text{ED}}\right|$. (b) Converged $\alpha$ for the TQS (markers) in (a) as a function of $t/U$, with dashed lines as a guide to the eye. (c) Histograms indicate the total relative frequencies, according to  \equref{eq:totalweight}, for the excitation classes $\mathcal{E}(\vec{s})$ from \equref{eq:labeldof}. From top to bottom, these refer to the insulating, critical and metallic regions indicated by the respective gray boxes in (a).  (d) HF-TQS results (markers) for the momentum-resolved fermionic bilinears $\mathcal{N}^j_{\vec{k}}$, defined in \equref{eq:observablen},  in comparison to those obtained from HF (dashed lines) for $N_{e}=12$.} \label{fig:bases}
\end{figure*}

\subsection{HF basis}
We first discuss the results using the HF basis. The solid gray lines in \figref{fig:scaling}a show the deviation of the HF ground state energy from that obtained via ED for the system sizes $N_e$ where the latter is feasible. As expected, the corrections exhibit higher magnitude near the metal-insulator transition (gray region). In the metallic regime ($t/U > 0.10$), the corrections decay more gradually, forming an extended tail. Contrastingly, on the insulating regime ($t/U < 0.05$), corrections decrease rapidly as $N_{e}$ is increased. To simultaneously display the performance of the Transformer-corrected ansatz (\ref{eq:tqsmod}) using the HF basis, which we refer to as HF-TQS from now on, the colored markers in \figref{fig:scaling}a show the deviations of HF from the HF-TQS ground-state energy. The fact that they are very close to the deviation of HF to ED for all parameters demonstrates the expressivity and convergence of our approach; this can also be more explicitly seen in the inset that directly shows the difference in ground-state energy between ED and HF-TQS. What is more, the HF-TQS ansatz can be applied to larger system sizes not accessible in ED (markers without gray line) which shows a behavior consistent with the values at smaller $N_e$ and still significant corrections to HF showing that admixing states in \equref{eq:tqsmod} beyond the Slater determinant $\ket{\text{RS}}$ provides a better approximation to the ground state. 

This behavior of the relevance of the RS can also be conveniently seen from the parameter $\alpha$ in \figref{fig:scaling}b. Away from the critical region, $\alpha$ approaches $1$ signaling that HF becomes an increasingly accurate approximation while, within the critical region, we find $\alpha\approx \sqrt{1-\alpha^{2}}\approx 0.7 \approx 1/\sqrt{2}$, indicating that only about half of the ground state or half of its energy [cf.~\equref{eq:expenergy}] is described by the HF state. 
The method's performance in this critical region is primarily limited by the total number of uniquely sampled states $n_{U}$ in \figref{fig:generalmethodology}c since, as we will see in more detail below, a larger number of states are needed in the critical region. When $n_{U}$ is sufficiently large to represent a substantial portion---or even the entirety---of the Hilbert space $\mathcal{H}$, the TQS converges with good accuracy. For all system sizes $N_{e}$, we fixed $n_{U}=4\times10^{3}$, which is sufficient to completely cover $\mathcal{H}$ for $N_{e}\le 12$.

Nonetheless, for most coupling parameters (i.e., not very close to the critical region), the HF-TQS can take crucial advantage of the physics-informed HF basis choice and efficiently capture the ground state at system sizes much larger than $N_e=12$.
This is demonstrated in \figref{fig:scaling}c, where we show the convergence of the ground state energy per site for $N_{e}=30$ in the metallic regime ($t/U=0.16$), revealing that the Transformer properly captures the non-product corrections to the HF state. In fact, we can see that the converged Transformer only ends up having to sample $n_{u}^{f}=1812$ distinct states (out of the $\approx 10^9$ total states). Before analyzing this in more detail, we point out that
the learning rate $\lambda_{\alpha_{0}}$ for optimizing $\alpha_{0}$ was set to a fixed value, such that the training dynamics is dominated by the one of $\vec{\theta}$ (\figref{fig:scaling}d). While alternative learning rate scheduling strategies could be proposed, they should be done with care. Specifically, we observed that low values of  $\lambda_{\alpha_{0}}$ can cause the optimization of $\vec{\theta}$, according  to \equref{eq:gradtheta}, to become trapped in local minima, particularly near to the phase transition.

\subsection{Other bases}
\label{subsec:otherbases}
We next compare the performance when using the HF basis with the chiral basis (see \secref{ModelHamiltonian}), with associated RS expected to provide a good approximation to the ground state for small $t/U$. To this end, we show in \figref{fig:bases}a the deviation of the variational ground state energy from HF (main panel) and ED (inset) for these bases choices. We see that the chiral and HF bases both provide accurate representations of the ground state across the entire phase diagram demonstrating again that the method is not intrinsically biased to being close to the RS, which, for the chiral basis, is not a good approximation for the ground state away from $t/U \rightarrow 0$; this is also confirmed by the behavior of the respective $\alpha$ shown in \figref{fig:bases}b: for the HF basis, it only dips significantly below $1$ in the critical region, where non-product-state corrections are crucial, while dropping to zero for increasing $t/U$ in the chiral basis.

To analyze the performance of our ansatz further, in \figref{fig:bases}(a,b), we also show results using the band basis [$U_{\vec{k}}=\mathbbm{1}$ in \equref{eq:basistransfo}], and choose a fully filled band (e.g., $a=-$) as RS which, importantly, is not close to the ground state for any $t/U$---not even in the non-interacting limit [as can be seen in \equref{eq:hamiltonian}, the band occupation has to change with momentum for $U=0$]. In line with these expectations, we find $\alpha \ll 1$ in the entire phase diagram. Nonetheless, the expressivity of the Transformer in the ansatz (\ref{eq:tqsmod}) allows to approximate the ground-state energy better than HF; it is not quite as good as in the HF or chiral basis which seems natural since the RS does not have any simple relation to the ground state in any part of the phase diagram. Thus, representing it and sampling from it is generically expected to be more challenging than in physics-informed bases. We checked that, for larger $t/U$, the transformer converges to the exact ground state energy also in the band basis as the asymptotic ground state is just one of the basis states (see \appref{sec:vmc}).

Additional important details about the wavefunction and sampling efficiency in the different bases can be revealed by studying the contributions of the various basis states. To group them, we recall that each $\ket{\vec{s}}$ in \equref{eq:tqsmod} is labeled by $\vec{s} = (s_1,\dots s_{N_{\vec{k}}})$, $s_k \in \{0,1\}$, and with the convention $\ket{\text{RS}} = \ket{(1,1,\dots,1)}$ it makes sense to use the number of ``excitations'' or ``flips'' 
\begin{equation}
\mathcal{E}(\vec{s}) = \sum_{k=1}^{N_{\vec{k}}} \left(1-s_{k}\right)
\label{eq:labeldof}
\end{equation} 
relative to the RS \footnote{For $N_{e}=6$, for example, states like $\ket{111110}$ and $\ket{111101}$ belong to the class with $\mathcal{E} =1$, i.e., with a single excitation above the RS}; in the case of the HF basis, these are in one-to-one correspondence to the particle-hole pairs described by the mean-field Hamiltonian (\ref{eq:unitarytransf}). The statistical weight of these groups can be quantified through their total relative frequency
\begin{equation}
R^{j}=\sum_{\vec{s}|\mathcal{E}(\vec{s})=j}r(\vec{s}) \quad \text{for} \quad j=1,\dots N_{e},
\label{eq:totalweight}
\end{equation}
where $r(\vec{s})$ is defined in \equref{eq:averages} and we excluded the RS, which is the only state corresponding to $j=0$ but has zero contribution to the Transformer by construction.

The histograms in \figref{fig:bases}c show these quantities for the three respective values of $t/U$ indicated in \figref{fig:bases}a using gray boxes. The chiral and band bases are clearly affected by the curse of dimensionality as all $R^{j}$ have significant contributions away from the insulating regime. In contrast, the ground state representation in the HF basis for both the insulating and metallic parameter range is dominated by low-order excitations relative to the HF state, as expected from \equref{eq:unitarytransf}. This behavior enables accurate calculations for $N_{e}\ge 16$ in \figref{fig:scaling}, both in the metallic and insulating regimes, in spite of the non-sparse nature of the Hamiltonian. 

Unlike coupled cluster methods in quantum chemistry, for example, the Transformer independently selects the most important excitation classes. This can be seen particularly from the HF-basis histogram close to the phase transition ($t/U=0.09$), as an increasing number of higher order excitations starts contributing to the ground-state energy. The number of accessible classes is then limited by only two factors: the total number of unique partial strings $n_{U}$ allowed in the batch-autoregressive sampler (\figref{fig:generalmethodology}b), and the Transformer's expressiveness, which is primarily controlled by $d_{\text{emb}}$, $N_{\text{h}}$ and $N_{\text{enc}}$ \cite{sanfordRepresentationalStrengthsLimitations2023} (see \figref{fig:generalmethodology}b and  \appref{sec:vmc}). Interestingly, though, we see in \figref{fig:bases}c that even close to the phase transition, the HF basis clearly benefits more from importance sampling than the other bases.

\begin{figure}[t]
    \centering
     \includegraphics[width=0.49\textwidth]{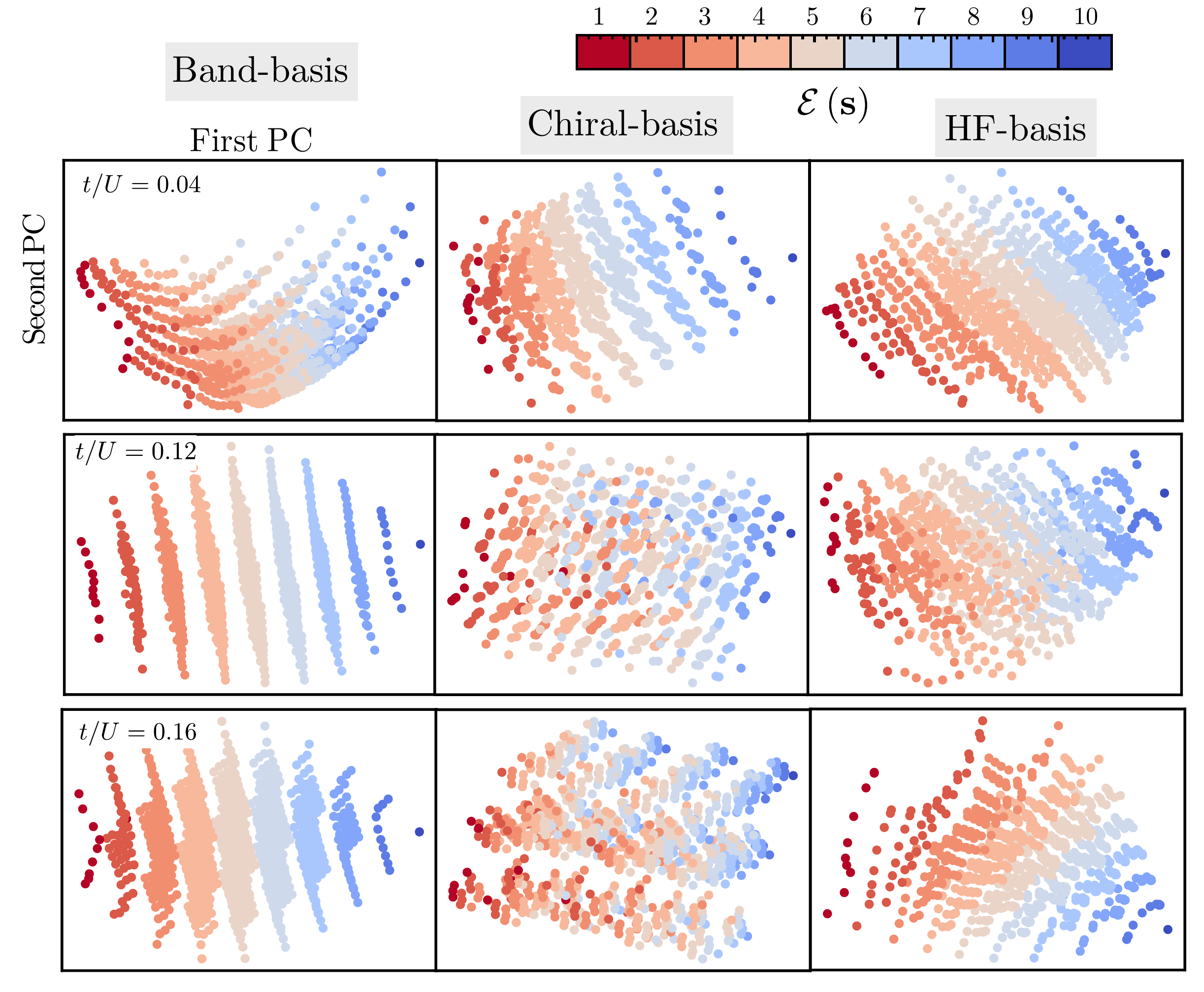}
    \caption{\textbf{Visualization of the Transformer's latent space.}  Results are shown for $N_{e}=10$ electrons at different values of $t/U$ for the band, chiral and HF bases. Each point represents a basis state $\vec{s}$, which is colored according to the class label $\mathcal{E}(\vec{s})$ [cf.~\equref{eq:labeldof}], and has been obtained by projecting the respective latent space features $\vec{H}(\vec{s})$ onto two dimensions using PCA. All  simulations use embedding dimension $d_{\text{emb}}=300$ with single attention head and decoder layer $N_{\text{h}}=N_{\text{dec}}=1$.}
    \label{fig:pcalatent}
\end{figure}

\subsection{Observables}
\label{subsec:obs}
Apart from the ground-state energy of \equref{eq:hamiltonian}, we can naturally estimate other observables, such as the momentum-resolved fermionic bilinears, 
\begin{equation}
    \mathcal{N}^j_{\vec{k}} = \bar{d}^{\,\dagger}_{\vec{k}} \sigma_j \bar{d}^{\,\pdagger}_{\vec{k}}, \quad j=x,y,z,
    \label{eq:observablen}
\end{equation}
where $\bar{d}_{\vec{k}}$ are the fermionic operators in the chiral basis. In \figref{fig:bases}d, we show their expectation values within HF and HF-TQS in the critical region ($t/U=0.09$). As the dispersion involves [first term in \equref{eq:hamiltonian}] $\sigma_x$ in the chiral basis, it is natural to recover the $\cos$-like shape in $\langle \mathcal{N}^x_{\vec{k}}\rangle $.
Most importantly, $\braket{\mathcal{N}^z_{\vec{k}}}$, which describes the symmetry breaking in the insulating regime, is sizeable in HF, showing that the system is already in the symmetry-broken, insulating regime. However, the additional quantum corrections from our HF-TQS approach lead to a much smaller almost vanishing $\braket{\mathcal{N}_{\vec{k}}^{z}}\approx 0$. This is in line with general expectations that HF overestimates the tendency to order.
Moreover, corrections to $\mathcal{N}^{y}_{\vec{k}}$ are more pronounced near the points where the kinetic term in \equref{eq:hamiltonian} changes sign (vertical blue lines in the plot). In combination with the fact that the deviations between HF and HF-TQS are much less pronounced away from the critical region (see \appref{sec:vmc}), these results demonstrate that the value of the parameter $\alpha$ ($\approx 0.76$ at $t/U=0.09$) also serves as an indicator of expected deviations from HF predictions for other physical observables.

\subsection{Hidden representation}
Finally, we investigate the influence of the three different bases on the Transformer's latent space by projecting the high-dimensional parametrization of $q_{\vec{\theta}}(\vec{s})$ onto low-dimensional spaces using principal component analysis (PCA) \cite{hotelling1933analysis, pearson1901liii, Huang2022Jul}. We apply this method to the set of vectors $\{\vec{H}(\vec{s})=\sum_{j}^{N_{e}} \vec{h}_{j}(\vec{s})| \forall\vec{s}\neq\text{RS}\}$ \cite{viterittiTransformerWaveFunction2024}, which are obtained at the output of the Transformer's $n_{\text{enc}}$ layers (see \figref{fig:generalmethodology}b). For visual clarity, we focus on $N_{e}=10$ electrons. Figure \ref{fig:pcalatent} shows the first and second components of PCA for all bases at $t/U=0.04$ (insulator), $t/U=0.12$ (close to critical region) and $t/U=0.16$ (metal). To first compare the two natural, energetically-motivated bases---the HF and chiral basis---we see that the states are indeed approximately ordered based on the classes defined via \equref{eq:labeldof} in the regimes where they are expected to be natural choices, i.e., for all $t/U$ (small $t/U$) for the HF (chiral) basis. This illustrates that the physical motivation for choosing these respective bases is not only visible in the histograms in \figref{fig:bases}b and the sampling efficiency but also ``learned'' by the Transformer's hidden representation. While some clear structure also emerges for the band basis, we emphasize that the labels $\mathcal{E}(\vec{s})$ do not directly translate to the energetics of the states: as discussed above, the RS is never close to the ground states in any regime, such that the number of excitations $\mathcal{E}$ above it also does not present clear energetic relevance either \footnote{Only for large $t/U$ does a related quantity, the excitations away from the product ground state, that can be defined in this basis become relevant.}.
The Transformer appears unable to uncover any additional emergent structure, which is likely related to the poor performance of the band basis, as shown in \figref{fig:bases}a. 

\section{Discussion}
We have introduced and demonstrated a modified transformer-based variational description of the ground state of a many-body Hamiltonian, which is based on first choosing an energetically motivated basis $\{\ket{\text{RS}},\ket{\vec{s}}\}$, see \equref{eq:tqsmod} and \figref{fig:generalmethodology}a. We showed that HF provides a very natural and general route towards finding such a basis since the associated mean-field Hamiltonian (\ref{eq:unitarytransf}) encodes an approximate energetic hierarchy of the states. As a second example, we used a basis defined in the strong-coupling limit. Overall, our approach has the following advantages: (i) there is a single parameter, $\alpha$, which quantifies how close the (variational representation of the) ground state is to $\ket{\text{RS}}$; for instance, for the HF basis, this would be the mean-field-theory prediction, i.e., the Slater determinant closest to the true ground state; (ii) except for right at the critical point, the HF basis is found to be particularly useful for improving the sampling efficiency since only a small subset of the exponentially large basis states contribute. This is expected based on general energetic reasoning and most directly visible in the histograms in \figref{fig:bases}c. Finally, (iii) the physical nature of these bases also allows for a clear interpretation of the different contributions, e.g., as excitations on top of the RS, which we also recover in the transformer’s hidden representation (see \figref{fig:pcalatent}). 

Several directions can be addressed in future work. First, applying this methodology to different Hamiltonians and deep learning architectures is a natural next step to see more generally under which conditions only a small subset of all basis states is required for the ground state in metallic and insulating regimes. 
From a methodological perspective, efficiency improvements could be achieved through the usage of modified stochastic reconfiguration techniques for the optimization of the network parameters \cite{chenEmpoweringDeepNeural2024a, Rende2024Aug}, and with the incorporation of symmetries in the HF-based ansatz \cite{Bao2024Jul, Pescia2024Jul}. For systems with non-sparse Hamiltonians like our current model, the implementation of the recently proposed GPU-optimized batch auto-regressive sampling without replacement \cite{Malyshev2024Aug} should also be beneficial. Furthermore, our approach could be used to test the validity and accuracy of different effective theories by using them as $\hat{H}_0$ in \figref{fig:generalmethodology}a to define the computational basis.

\begin{acknowledgments}

M.S.S. thanks P.~Wilhelm for discussions and previous collaborations. J.A.S. also acknowledges discussions with Y.-H.~Zhang, S.~Banerjee, L.~Pupim, V.~Dantas, P.~Wilhelm, M.~Mühlbauer,  and M.~Medvidović.
\end{acknowledgments}

\bibliography{draft_Refs}

\onecolumngrid

\begin{appendix}

\section{Fermionic Model}
\label{sec:fm}
\subsection{General basis}

The Hamiltonian \eqref{eq:hamiltonian}  can also be naturally represented in a general basis as 
\begin{equation}
    \hat{H}= t\sum_{\vec{k}\in\text{BZ}} \cos(\vec{k}) \, \bar{d}^\dagger_{\vec{k}, \alpha} \mathcal{P}_{\alpha, \beta}\left(\vec{k}\right) \bar{d}^\pdagger_{\vec{k},\beta} + U \sum_{\vec{q} \in \mathbbm{R}} V(\vec{q}) \rho_{\vec{q}} \rho_{-\vec{q}}, \label{eq:hamiltonian2} 
\end{equation}
with the density operator 
\begin{equation}
    \rho_{\vec{q}} = \sum_{k\in\text{BZ}} \left(\bar{d}^\dagger_{\text{BZ}(\vec{k}+\vec{q}), \alpha} \mathcal{F}_{\alpha, \beta}\left(\vec{k}, q\right) \bar{d}^\pdagger_{\vec{k}, \beta} - \sum_{\vec{G}\in\text{RL}} \delta_{q,G} f_{1}(\vec{k},\vec{G})\right),
\label{eq:densityoperator0}
\end{equation}
where BZ indicates the Brillouin Zone defined as $[-\pi, \dots,\pi-2\pi/N_{\vec{k}}]$. The form factors are given by
\begin{equation}
    \mathcal{P}_{\alpha,\beta}(\vec{k}) = \left(U_{\vec{k}}^\dagger \sigma_x U_{\vec{k}}^\pdagger\right)_{\alpha,\beta}
    \label{eq:formfactor1}
\end{equation}
and
\begin{equation}
\mathcal{F}_{\alpha,\beta}(\vec{k},\vec{q}) = \sum_{\gamma =\pm} \left(U_{\vec{k}}^\dagger\right)_{\alpha, \gamma}\left[ f_1(\vec{k},\vec{q}) + i \gamma f_{2}(\vec{k},\vec{q}) \right] \left(U_{\vec{k}}^\pdagger\right)_{\gamma,\beta }.
\label{eq:formfactor22}
\end{equation}
The expressions $f_{j}\left(\vec{k},\vec{q}\right)$  in eq. \equref{eq:formfactor22} can be further determined based on desired symmetries for the model. Since $\rho_{\vec{q}}$ is the Fourier transform of the density operator, it needs to obey $\rho^\dagger_{\vec{q}} = \rho_{-\vec{q}}$. We impose $C_{2z}$ invariance, which leads to the following constraints on the form factors 
\begin{equation}
\begin{aligned}
    f_{1}(\vec{k},\vec{q}) &= f_{1}(\text{BZ}(\vec{k}+\vec{q}),-\vec{q}) \\
    f_{2}(\vec{k},\vec{q}) &= -f_{2}(\text{BZ}(\vec{k}+\vec{q}),-\vec{q}) \\
    f_{j}(\vec{k},\vec{q}) &= f_{j}(-\vec{k},-\vec{q}).\label{hermiticityandc2z}
\end{aligned}
\end{equation}
In turn, these imply that
\begin{equation}
\begin{aligned}
    f_{1}(\vec{k},\vec{G}) &= f_{1}(-\vec{k},\vec{G}), \\ 
    f_{2}(\vec{k},\vec{G}) &= -f_{2}(-\vec{k},\vec{G}), \quad  \forall \,  \vec{G} \in \text{RL}.
    \label{eq:formfactor2}
\end{aligned}
\end{equation}
A possible choice that satisfies Eqs.~\eqref{hermiticityandc2z} and \eqref{eq:formfactor2} is given by
\begin{equation}
f_{j}\left(k,q\right) = 
\begin{cases}
1,  \;\text{for } j=1\\
 \beta \sin \left(k\right) \left(\sin \left(q\right) +\sin \left(k+q\right)\right), \; \text{for } j=2. \\
\end{cases}
\label{eq:formfactors}
\end{equation}
Here, $\beta$ mediates the strength difference between both form factors. Without loss of generality, we set $\beta=0.9$. 

The chiral (interacting limit) and band (kinetic limit) bases can be retrieved from Eqs.~\eqref{eq:hamiltonian2}-\eqref{eq:formfactor22} as follows. For $t/U\rightarrow 0$, we can consider the states 
\begin{equation}
    \ket{\pm} = \prod_{\vec{k}\in\text{BZ}} \bar{d}^\dagger_{\vec{k},\pm} \ket{0}, \label{eq:chiralbasis}
\end{equation}
where $\bar{d}^\dagger_{\vec{k},\pm}$ are the creation operators defined with $U_{\vec{k}}=\mathbb{I}$ in \eqref{eq:hamiltonian2}. The Hamiltonian takes a positive semi-definite form $\hat{H}\propto \sum_{\vec{q} \in \mathbbm{R}} V(\vec{q}) \rho_{\vec{q}}^{\pdagger} \rho^\dagger_{\vec{q}}$. Given that the states \eqref{eq:chiralbasis} obey $\rho_{\vec{q}} \ket{\psi_\pm} = 0$, we conclude that they represent the ground state of the system, and, thus, the model becomes exactly solvable in this limit. 

Similarly, the \textit{band}-basis, defined with $
    U_{\vec{k}}= \frac{1}{\sqrt{2}} \begin{pmatrix} 1 & -i \\ 1 & i \end{pmatrix},$ in \equref{eq:hamiltonian2}, acquires the diagonal form $\hat{H}\propto\sum_{\vec{k} \in \mathrm{BZ}} \cos (\vec{k}) d_{\vec{k}}^{\dagger} \sigma_z d_{\vec{k}}^{\pdagger}=\sum_{\vec{k} \in \mathrm{BZ}} \sum_{j=1,2} \varepsilon_{\vec{k}, j} d_{\vec{k}, j}^{\dagger} d_{\vec{k}, j}$ in band space with $\varepsilon_{\vec{k}, j}=\left(-1\right)^{j}\cos\vec{k}$ for large $t/U$. If the total number of electrons $N_{e}$ is chosen such that $\{-\frac{\pi}{2},\frac{\pi}{2}\}\notin \text{BZ}$, the ground state is obtained by filling the lower band as
\begin{equation}
\left|\psi_{-}\right\rangle=\prod_{\vec{k}\in \mathrm{BZ}} d_{\vec{k},-}^{\dagger}|0\rangle \quad \text{where} \quad d_{\vec{k},-}^{\dagger}= \begin{cases}d_{\vec{k}, 1}^{\dagger} & |\vec{k}|<\frac{\pi}{2} \\ d_{\vec{k}, 0}^{\dagger} & |\vec{k}|>\frac{\pi}{2}\end{cases}.
\label{eq:exactbigt}
\end{equation}

For finite $t/U$, an optimal basis choice becomes non-trivial. Some intuition can be acquired when expressing the kinetic term in the chiral basis, which yields $\hat{H}=\sum_{\vec{k} \in \mathrm{BZ}} \cos (\vec{k}) \bar{d}_{\vec{k}}^{\dagger} \sigma_x \bar{d}_{\vec{k}}^{\pdagger}$. This structure indicates that within second order perturbation theory, other states $\vec{s}$ will be admixed to \eqref{eq:chiralbasis}, where some $d_{\vec{k}\pm}^{\dagger}$ will be replaced with $d_{\vec{k}\mp}^{\dagger}$. Despite this mixing, the $\ket{\psi_\pm}$ will remain degenerate since the Hamiltonian is invariant under the $\vec{k}$-local anti-unitary symmetry described by the anti-unitary operator $PT$ with
$PTd_{\vec{k}} (PT)^\dagger = \sigma_x d_{\vec{k}}$
and $PT\ket{\psi_+} \propto \ket{\psi_-}$, where $T$ indicates the time reversal symmetry operator.

\subsection{Expectation values on the chiral-basis}

For the evaluation of the energy expectation value, it is instructive to start with a fixed basis.  We choose the chiral-basis, and then generalize it to an arbitrary basis. For notational simplicity, we omit explicit mention of the $\ket{\text{RS}}$ state  removal in the following considerations, as they remain valid whether or not we include it. 
The variational ansatz can be written as 
\begin{equation}
    \ket{\psi_{\vec{\theta}}} = \sum_{\vec{s}\in [+,-]^{N_{e}}} \psi_{\vec{\theta}}(\vec{s}) \prod_{k} \bar{d}^\dagger_{k,(\vec{s})_k} \ket{0}. \label{FirstVariationalAnsatz}
\end{equation}
For $t/U=0$ the ground state has $\psi_{\vec{\theta}}(\vec{s})=0$ except for $\vec{s}=(+,+,+,\dots ,+)$ (or all $-$). Finite $t$ will admix other $\vec{s}$ configurations.

To calculate the expectation value of $\hat{H}$, we need the matrix elements 
\begin{equation}
    H_{\vec{s}\vec{s}'} := \braket{\vec{s}|H|\vec{s}'},
\end{equation}
where $\ket{\vec{s}}$ are the basis states.  Without loss of generality, we consider the band mapping $s_{i}=0,1 \rightarrow -1, 1 (-,+)$ to what follows. Additionally, we introduce the notation 
\begin{equation}
[\boldsymbol{s}]_{k}=\begin{cases}
\left(s\right)_{k^{\prime}} \quad \text{for} \; k\neq k^{\prime}\\
-\left(s\right)_{k}
\quad \text{for} \; k= k^{\prime}\\
\end{cases}
\label{eq:hoppings}
\end{equation}
to indicate a state which is equal to $\vec{s}$ apart from the hopping of one electron to another band at the $k$-position. Let us begin with the non-interacting (kinetic) part $\hat{H}_0$  of the Hamiltonian:
\begin{equation}
    \braket{\vec{s}|\hat{H}_0|\vec{s}'} = t\sum_{\vec{k}}\cos(\vec{k}) \delta_{[\vec{s}]_{\vec{k}}\vec{s}'}.\label{SimpleHopping}
\end{equation}
The associated ``local Hamiltonian'' then simply reads as
\begin{equation}
    \hat{H}^{0}_{\text{loc}}(\vec{s}) = t\sum_{\vec{k}}\cos(\vec{k}) \frac{\psi_{\vec{\theta}}([\vec{s}]_k)}{\psi_{\vec{\theta}}(\vec{s})}.
\end{equation}
according to \equref{eq:hloc}. If we only have this non-interacting part (or asymptotically in the limit $|t/U|\gg 1$), we will therefore have
\begin{equation}
    \psi_{\vec{\theta}}(\vec{s}) = \prod_k  \frac{1}{\sqrt{2}}\left( \delta_{(\vec{s})_k,+} - \sign(t\cos k) \delta_{(\vec{s})_k,-} \right)
\end{equation}
in the ground state (with associated energy $E=-t\sum_{\vec{k}} |\cos \vec{k}|$). The momentum-resolved fermionic bilinears defined in \equref{eq:observablen} 
can be calculated on a similar manner. For the x-component, for example
\begin{equation}
    \braket{\vec{s}|\mathcal{N}^{x}_{\vec{k}}|\vec{s}'} = \delta_{[\vec{s}]_k,\vec{s}'}, \qquad \mathcal{N}^{x,\text{loc}}_k(\vec{s}) = \frac{\psi_{\vec{\theta}}([\vec{s}]_k)}{\psi_{\vec{\theta}}(\vec{s})}.
    \label{eq:expvalueNk}
\end{equation}

Now we focus on the interaction part, i.e., $\hat{H}_1 = \sum_{\vec{q} \in \mathbbm{R}} V(\vec{q}) \rho_{\vec{q}} \rho_{-\vec{q}}$. To compute the matrix elements $\braket{\vec{s}|H_1|\vec{s}'}$, we consider the expressions
\begin{equation}
    \braket{\vec{s}|\bar{d}^\dagger_{\vec{k}+\vec{q},p}\bar{d}^\pdagger_{\vec{k},p}|\vec{s}'} = \delta_{\vec{q}\in\text{RL}}\delta_{\vec{s},\vec{s}'}\delta_{(\vec{s})_k,p},
    \label{eq:decompositionchiral1}
\end{equation}
and
\begin{align}
         &\braket{\vec{s}|\bar{d}^\dagger_{\vec{k}+\vec{q},p}\bar{d}^\pdagger_{\vec{k},p}\bar{d}^\dagger_{\vec{k}'-\vec{q},p'}\bar{d}^\pdagger_{\vec{k}',p'}|\vec{s}'}   = \delta_{\vec{q}\in\text{RL}}\delta_{\vec{s},\vec{s}'}\delta_{(\vec{s})_k,p}\delta_{(\vec{s})_{k'},p'} + \nonumber \\ &  + \delta_{\vec{q}\notin\text{RL}} \delta_{\vec{k},\text{BZ}(\vec{k}'-\vec{q})} \delta_{(\vec{s}')_{k'},p'}\delta_{(\vec{s}')_{k'-q},-p'}\delta_{(\vec{s})_{k'-q},-p}\delta_{(\vec{s})_{k'},p}\left(\prod_{k_1\neq k',k'-q}\delta_{(\vec{s})_{k_1},(\vec{s}')_{k_1}}\right)\left[\delta_{p,p'}-\delta_{p,-p'}\right].
    \label{eq:decompositionchiral2}
\end{align}
Here, we introduced $\delta_{\vec{q}\in\text{RL}} := \sum_{\vec{G}\in\text{RL}} \delta_{\vec{q},\vec{G}}$ and $\delta_{\vec{q}\notin\text{RL}} := 1 -\delta_{\vec{q}\in\text{RL}}$. The fact that only a small subset of matrix elements is non-zero follows from the restriction to states without double occupancy. Sticking to the current approximation, we end up with the following form of the matrix elements
\begin{align}
    \braket{\vec{s}|\hat{H}_1|\vec{s}'} &= \delta_{\vec{s},\vec{s}'} h_1(\vec{s})
    +  \sum_{\vec{q}\in\mathbbm{R}\setminus \text{RL}}V(\vec{q}) \sum_{\vec{k}\in\text{BZ}}\sum_{p,p'} f^*_p(\vec{k},-\vec{q})f_{p'}(\vec{k},-\vec{q}) \left[\delta_{p,p'}-\delta_{p,-p'}\right]  \nonumber \\ &\quad \times\delta_{(\vec{s}')_{k'},p'}\delta_{(\vec{s}')_{k'-q},-p'}\delta_{(\vec{s})_{k'-q},-p}\delta_{(\vec{s})_{k'},p}\left(\prod_{k_1\neq k',k'-q}\delta_{(\vec{s})_{k_1},(\vec{s}')_{k_1}}\right),
\end{align}
where $f_p\left(\vec{k},\vec{q}\right) = f_1\left(\vec{k},\vec{q}\right) + i p f_2\left(\vec{k},\vec{q}\right)$ and the diagonal matrix elements are given by
\begin{align}
    h_1(\vec{s}) &= \sum_{\vec{G}\in\text{RL}}V(\vec{G}) \sum_{\vec{k},\vec{k}'} \Biggl[f_1(\vec{k},\vec{G})f_1(\vec{k}',-\vec{G}) - \sum_{r=\pm} f_{1}(\vec{k},r \vec{G}) \sum_p f_p(\vec{k}',-r\vec{G}) \delta_{(\vec{s})_{k'},p} \\
    &\qquad\quad + \sum_{p,p'} f_p(\vec{k},\vec{G})f_{p'}(\vec{k}',-\vec{G}) \delta_{(\vec{s})_{k},p}\delta_{(\vec{s})_{k'},p'} \Biggr].
\end{align}
As a simple consistency check, one can see from these expressions that $\braket{\vec{s}|\hat{H}_1|(p_0,p_0,\dots,p_0)}=0$ for any $\vec{s}$ and $p_0=\pm$, as it should be [recall \equref{eq:chiralbasis}].

More generally, due to (i) the conditions imposed to the form factors in Eqs.~\eqref{hermiticityandc2z}-\eqref{eq:formfactor2},
and that (ii) double occupancy is  neglected, we can show that terms proportional to $\delta_{\vec{q}\in \text{RL}}$ do not contribute to the ground state energy, i.e., that $\hat{H}_{1}^{\vec{G}}=\sum_{\vec{G}\in \text{RL}}V(\vec{G})\rho^{\pdagger}_{\vec{G}}\rho_{\vec{G}}^{\dagger}=0$, or equivalently, that all matrix elements $\braket{\vec{s}|\hat{H}_{1}^{\vec{G}}|\vec{s}^{\prime}}=0$. From, 
\begin{equation}
\rho_{\vec{G}}=\sum_{\vec{k}\in \text{BZ}}\left(\sum_{p=\pm}\bar{d}^{\dagger}_{\vec{k}, p}f_{1}\left(\vec{k}, \vec{G}\right)\bar{d}^{\pdagger}_{\vec{k}, p}-f_{1}\left(\vec{k},\vec{G}\right)\right)
\label{eq:density}
\end{equation}
we start by decomposing the matrix elements in 
\begin{equation}
\begin{aligned}
\langle\boldsymbol{s}| \hat{H}_1^{\vec{G}}\left|\boldsymbol{s}^{\prime}\right\rangle= & \sum_{\vec{G} \in \text{RL}} V(\vec{G}) \sum_{\vec{k}, \vec{k}^{\prime} \in \mathrm{BZ}} f_1(\vec{k}, \vec{G}) f_1\left(\vec{k}^{\prime}, \vec{G}\right) \\
& \left(-2 \sum_{p= \pm}\langle\boldsymbol{s}| \bar{d}_{\vec{k}, p}^{\dagger} \bar{d}_{\vec{k}, p}^{\pdagger}\left|\boldsymbol{s}^{\prime}\right\rangle+\sum_{p, p^{\prime}= \pm}\langle\boldsymbol{s}| \bar{d}_{\vec{k}, p}^{\dagger} \bar{d}_{\vec{k}, p} \bar{d}_{\vec{k}^{\prime}, p^{\prime}}^{\dagger} \bar{d}_{\vec{k}^{\prime}, p^{\prime}}\left|\boldsymbol{s}^{\prime}\right\rangle+\left\langle\boldsymbol{s} \mid \boldsymbol{s}^{\prime}\right\rangle\right),
\end{aligned}
\end{equation}
where
\begin{equation}
\begin{aligned}
\langle\boldsymbol{s}| \bar{d}_{\vec{k}, p}^{\dagger} \bar{d}_{\vec{k}, p}^{\pdagger}\left|\boldsymbol{s}^{\prime}\right\rangle & =\delta_{\left(\boldsymbol{s}^{\prime}\right)_{k}, p^{\prime}} \delta_{(\boldsymbol{s})_k, p} \delta_{\boldsymbol{s},\boldsymbol{s}^{\prime}}=\delta_{(\boldsymbol{s})_k, p} \delta_{\boldsymbol{s}, \boldsymbol{s}^{\prime}} \\
\langle\boldsymbol{s}| \bar{d}_{\vec{k}, p}^{\dagger} \bar{d}_{\vec{k}, p}^{\pdagger} \bar{d}_{\vec{k}^{\prime}, p^{\prime}}^{\dagger} \bar{d}^{\pdagger}_{\vec{k}^{\prime}, p^{\prime}}\left|\boldsymbol{s}^{\prime}\right\rangle & =\delta_{\left(\boldsymbol{s}^{\prime}\right)_{k^{\prime}}, p^{\prime}} \delta_{(\boldsymbol{s})_k, p} \delta_{\boldsymbol{s}, \boldsymbol{s}^{\prime}}
\end{aligned}
\end{equation}
leads to 
\begin{equation}
\begin{aligned}
\langle\boldsymbol{s}| \hat{H}_1^{\vec{G}}\left|\boldsymbol{s}^{\prime}\right\rangle=\delta_{\vec{s}, \vec{s}^{\prime}} & \sum_{\vec{G} \in \text{RL}} V(\vec{G}) \sum_{\vec{k}, \vec{k}^{\prime} \in \mathrm{BZ}} f_1(\vec{k}, \vec{G}) f_1\left(\vec{k}^{\prime}, \vec{G}\right) \\
& \left(-2 \sum_{p= \pm} \delta_{(\boldsymbol{s})_k, p}+\sum_{p, p^{\prime}= \pm} \delta_{\left(\vec{s}^{\prime}\right)_{k^{\prime}}, p^{\prime}} \delta_{(\vec{s})_k, p}+1\right).
\end{aligned}
\end{equation}
The expression inside the parenthesis is zero for any combination of $\vec{s}$ and $\vec{s}^{\prime}$ if double occupancy is neglected. 
\subsection{Expectation values in a general basis}

When working in a general basis, as defined by the transformation \equref{eq:basistransfo} with $U_k$ determined from HF, the new variational ansatz will be given as
\begin{equation}
    \ket{\psi_{\vec{\theta}}} = \sum_{\vec{s}\in [+,-]^{N_{e}}} \psi_{\vec{\theta
    }}(\vec{s}) \prod_{k} \bar{d}^\dagger_{\vec{k},(\vec{s})_k} \ket{0}. \label{VariationalAnsatz}
\end{equation}
In this case, we need the modified matrix elements
\begin{equation}
    \hat{\bar{H}}_{\vec{s}\vec{s}'} :=\braket{\bar{\vec{s}}|\hat{H}|\bar{\vec{s}}'}, \qquad \ket{\bar{\vec{s}}} := \prod_{k} \bar{d}^\dagger_{\vec{k},(\vec{s})_k} \ket{0}, \label{MatrixElementsInBarredBasis}
\end{equation}
according to Eqs.~\eqref{eq:hamiltonian2}-\eqref{eq:formfactor22}. 
The computation of the matrix elements (\ref{MatrixElementsInBarredBasis}) closely parallels that of the previous subsection. For the non-interacting part, we get
\begin{equation}
    \braket{\vec{s}|\hat{H}_0|\vec{s}'} = t\sum_{\vec{k}}\cos(\vec{k}) \left[ \delta_{\vec{s},\vec{s}'}\sum_{\alpha=\pm} \mathcal{P}_{\alpha,\alpha}(\vec{k})\delta_{(\vec{s})_k,\alpha} + \delta_{[\vec{s}]_k,\vec{s}'} \sum_{\alpha=\pm}\mathcal{P}_{\alpha,-\alpha}(\vec{k}) \delta_{(\vec{s})_k,\alpha} \right].
    \label{eq:h0localgeneralbasis}
\end{equation}
which allows us to identify the local non-interacting energy as 
\begin{equation}
\begin{aligned}
\hat{H}^{0}_{\text{loc}}(\boldsymbol{s}) & =t \sum_{\vec{k}} \cos (\vec{k})\left[\mathcal{P}_{\alpha, \alpha}(\vec{k})+\mathcal{P}_{\alpha,-\alpha}(\vec{k}) \frac{\psi_{\vec{\theta}}\left([\boldsymbol{s}]_k\right)}{\psi_{\vec{\theta}}(\boldsymbol{s})}\right], \quad \text{with} \quad 
\alpha  =(\boldsymbol{s})_k.
\end{aligned}
\end{equation}

To write down the matrix elements $\braket{\vec{s}|\hat{H}_1|\vec{s}'}$ of the interacting part, we first compute the matrix elements
\begin{align}
    \braket{\vec{s}|\bar{d}^\dagger_{\vec{k}+\vec{q},\alpha}\bar{d}^\pdagger_{\vec{k},\beta}|\vec{s}'} &= \delta_{\vec{q}\in\text{RL}}\left[\delta_{\alpha,\beta}\delta_{\vec{s},\vec{s}'}\delta_{(\vec{s})_k,\alpha} + \delta_{\alpha,-\beta} \delta_{[\vec{s}]_k,\vec{s}'} \delta_{(\vec{s})_k,\alpha}  \right], \nonumber \\
    \braket{\vec{s}|\bar{d}^\dagger_{\vec{k}+\vec{q},\alpha}\bar{d}^\pdagger_{\vec{k},\beta}\bar{d}^\dagger_{\vec{k}'-\vec{q},\gamma}\bar{d}^\pdagger_{\vec{k}',\delta}|\vec{s}'}  \nonumber  \\ & \hspace{-10em} = \delta_{\vec{q}\in\text{RL}} \left[\delta_{\vec{k}\neq \vec{k}'}\delta_{(\vec{s}')_{k'},\delta}\delta_{(\vec{s})_{k'},\gamma}\delta_{(\vec{s}')_{k},\beta}\delta_{(\vec{s})_{k},\alpha} + \delta_{\vec{k},\vec{k}'} \delta_{\beta,\gamma} \delta_{(\vec{s})_{k},\alpha} \delta_{(\vec{s}')_{k},\delta} \right] \left(\prod_{k_1\neq k',k'-q}\delta_{(\vec{s})_{k_1},(\vec{s}')_{k_1}}\right)   \\  & \hspace{-10em}+ \delta_{\vec{q}\notin\text{RL}}  \delta_{\vec{k},\text{BZ}(\vec{k}'-\vec{q})} \delta_{(\vec{s}')_{k'},\delta}\delta_{(\vec{s}')_{k'-q},-\gamma}\delta_{(\vec{s})_{k'-q},-\beta}\delta_{(\vec{s})_{k'},\alpha}\left(\prod_{k_1\neq k',k'-q}\delta_{(\vec{s})_{k_1},(\vec{s}')_{k_1}}\right)\left[\delta_{\beta,\gamma}-\delta_{\beta,-\gamma}\right],\nonumber 
\end{align}
With this, the evaluation of $\braket{\vec{s}|\hat{H}_1|\vec{s}'}$ can be performed for any unitary basis according to \equref{eq:basistransfo}. As another consistency check we reproduce the Eqs.~\eqref{eq:decompositionchiral1}-\eqref{eq:decompositionchiral2} of the previous section in the limit $\alpha=\beta=p$ and $\gamma=\delta=p'$.

From this consideration, the local interacting contribution to the ground state energy is given by 
\begin{equation}
\hat{H}_{\text{loc}}^{ 1}= \sum_{\vec{q} \notin \mathrm{RL}} V(\vec{q}) \sum_{\vec{k} \in \mathrm{BZ}} \sum_{\delta \in\{+,-\}} \mathcal{F}_{\alpha, \beta}(\mathrm{BZ}(\vec{k}-\vec{q}), \vec{q})  \times\left[\mathcal{F}_{\beta, \delta}(\vec{k},-\vec{q}) \frac{\psi_{\vec{\theta}}\left([s]_{\vec{k} \rightarrow \delta}\right)}{\psi_{\vec{\theta}}(s)}-\mathcal{F}_{-\beta, \delta}(\vec{k},-\vec{q}) \frac{\psi_{\vec{\theta}}\left([\boldsymbol{s}]_{k \rightarrow \delta ; \mathrm{BZ}(k-q)}\right)}{\psi_{\vec{\theta}}(\boldsymbol{s})}\right], 
\end{equation}
with $\alpha=  (\boldsymbol{s})_k$,  $\beta=-(\vec{s})_{\mathrm{BZ}(k-q)}$. According to the notation introduced in \equref{eq:hoppings}, we also define  
\begin{equation}
[\boldsymbol{s}]_{k \rightarrow \delta}=\begin{cases}
\left(s\right)_{k^{\prime}} \quad \text{for} \; k\neq k^{\prime}\\
\delta 
\quad \text{for} \; k= k^{\prime}\\
\end{cases},
\end{equation}
and
\begin{equation}
[\boldsymbol{s}]_{k \rightarrow \delta ; l}=\begin{cases}
\left(s\right)_{k^{\prime}} \quad \text{for} \; k\neq k^{\prime}\; \text{and}\;k^{\prime} \neq l\\
\delta 
\quad\text{for} \; k= k^{\prime}\\
-\left(s\right)_{l}
\quad \text{for} \; k= l\\
\end{cases}.
\end{equation}
Finally, expression \eqref{eq:expvalueNk} reads now as 
\begin{equation}
\begin{aligned}
\mathcal{N}_{\vec{k}}^{\text{loc}}(\boldsymbol{s}) & =\mathcal{P}_{\alpha, \alpha}(\vec{k})+\mathcal{P}_{\alpha,-\alpha}(\vec{k}) \frac{\psi_{\vec{\theta}}\left([\boldsymbol{s}]_k\right)}{\psi_{\vec{\theta}}(\boldsymbol{s})} \\
\alpha & =(\boldsymbol{s})_k.
\end{aligned}
\end{equation}

\section{Hartree-Fock implementation for the fermionic model}\label{HFAppendix}

Now we show how to obtain the self-consistent HF equations to the Hamiltonian \eqref{eq:hamiltonian}, following Ref. \cite{Christos2022Apr}. Starting on the chiral basis, without loss of generality, we note that the kinetic term is  already equivalent to its mean-field expression
\begin{align}
	H_0^{\text{HF}} &= \sum_{\vec{k}\in\text{BZ}} d^\dagger_{\vec{k}} \widetilde{h_{\vec{k}}} d^\pdagger_{\vec{k}}, \label{HFKinetic}
	\quad \text{with} \quad \widetilde{h_{\vec{k}}} = t\cos(\vec{k}) \sigma^\pdagger_x.
\end{align}
For the interacting term we start from
\begin{align}
	\hat{H}_1 &= \sum_{\vec{q}\notin\text{RL}} V(\vec{q}) \rho_{\vec{q}} \rho_{-\vec{q}}  \nonumber \\
	&= \sum_{\vec{q}\notin\text{RL}} V(\vec{q})  \sum_{\vec{k},\vec{k}'\in\text{BZ}}
	d^\dagger_{\text{BZ}(\vec{k}+\vec{q}),\alpha}  d^\pdagger_{\vec{k},\beta}
	d^\dagger_{\text{BZ}(\vec{k}'-\vec{q}),\gamma}  d^\pdagger_{\vec{k}',\delta} \mathcal{F}_{\alpha,\beta}(\vec{k},\vec{q})^\pdagger \mathcal{F}_{\gamma,\delta}(\vec{k}',-\vec{q})^\pdagger. 
\end{align}
This can be brought to normal order, by using the fermionic anti commutation relation $\{d^\pdagger_{\vec{k}\beta},
d^\dagger_{\text{BZ}(\vec{k}'-\vec{q}),\gamma}\} = \delta^\pdagger_{\vec{k}, \text{BZ}(\vec{k}'-\vec{q})}\delta^\pdagger_{\beta, \gamma}$, as 
\begin{align}
	\hat{H}_{1} = &-\sum_{\vec{q}\notin\text{RL}} V(\vec{q}) \sum_{\vec{k},\vec{k}'\in\text{BZ}}
	d^\dagger_{\text{BZ}(\vec{k}+\vec{q}),\alpha} d^\dagger_{\text{BZ}(\vec{k}'-\vec{q}),\gamma} d^\pdagger_{\vec{k},\beta} d^\pdagger_{\vec{k}',\delta} \mathcal{F}_{\alpha\beta}(\vec{k},\vec{q})^\pdagger \mathcal{F}_{\gamma\delta}(\vec{k}',-\vec{q})^\pdagger \;+\nonumber \\
	&+ \sum_{\vec{q}\notin\text{RL}} V(\vec{q}) \sum_{\vec{k},\vec{k}'\in\text{BZ}}
	d^\dagger_{\text{BZ}(\vec{k}+\vec{q}),\alpha} d^\pdagger_{\vec{k}'\delta} \mathcal{F}_{\alpha\beta}(\vec{k},\vec{q})^\pdagger \mathcal{F}_{\gamma\delta}(\vec{k}',-\vec{q})^\pdagger \delta^\pdagger_{\vec{k}, \text{BZ}(\vec{k}'-\vec{q})}\delta^\pdagger_{\beta, \gamma} \label{eq:NormalOrderRemainingSingleBody}.
\end{align}
The single-body term can be further simplified by first shifting the index $\vec{k} \rightarrow \text{BZ}(\vec{k} - \vec{q})$ and $\vec{q} \rightarrow -\vec{q}$,
\begin{align}
	\hat{H}_1^{\text{single}} &= \sum_{\vec{q}\notin\text{RL}} V(\vec{q}) \sum_{\vec{k}, \vec{k}'\in\text{BZ}}
	d^\dagger_{\text{BZ}(\text{BZ}(\vec{k}-\vec{q})+\vec{q}),\alpha} d^\pdagger_{\vec{k}',\delta} \mathcal{F}_{\alpha\beta}(\text{BZ}(\vec{k}-\vec{q}),\vec{q})^\pdagger 	\mathcal{F}_{\gamma\delta}(\vec{k}',-\vec{q})^\pdagger \delta^\pdagger_{\text{BZ}(\vec{k}-\vec{q}), \text{BZ}(\vec{k}'-\vec{q})}\delta^\pdagger_{\beta, \gamma} \nonumber \\
	&= \sum_{\vec{q}\notin\text{RL}} V(\vec{q}) \sum_{\vec{k}\in\text{BZ}}
	d^\dagger_{\vec{k}\alpha} d^\pdagger_{\vec{k}\delta} \mathcal{F}_{\alpha\beta}^\dagger(\vec{k},-\vec{q}) \mathcal{F}_{\beta\delta}^\pdagger(\vec{k},-\vec{q})^\pdagger \nonumber \\
	&= \sum_{\vec{q}\notin\text{RL}} V(\vec{q}) \sum_{\vec{k}\in\text{BZ}}
	d^\dagger_{\vec{k}} \mathcal{F}^\dagger(\vec{k},\vec{q}) \mathcal{F}(\vec{k},\vec{q}) d^\pdagger_{\vec{k}} \nonumber \\
	&= \sum_{\vec{k}\in\text{BZ}} 
	d^\dagger_{\vec{k}}
	h_1^{\text{single}}
	d^\pdagger_{\vec{k}},
\end{align}
where 
\begin{align}
	h_1^{\text{single}} &= \sum_{\vec{q}\notin\text{RL}}
	V(\vec{q}) \mathcal{F}^\dagger(\vec{k},\vec{q}) \mathcal{F}(\vec{k},\vec{q})^\pdagger.
	\label{ResultingH1Single}
\end{align}
This new term is included in the single body term in \equref{HFKinetic} by replacing $\widetilde{h_{\vec{k}}} \rightarrow h_{\vec{k}} = \widetilde{h_{\vec{k}}} + h_{1}^{\text{single}}$.  

Proceeding to the first term in \equref{eq:NormalOrderRemainingSingleBody}
we first shift the index $\vec{k}' \rightarrow \text{BZ}(\vec{k}'+\vec{q})$ and rearrange the annihilation operators as
\begin{align}
	\hat{H}_1^{\text{eff}} &= -\sum_{\vec{q}\notin\text{RL}} V(\vec{q}) \sum_{\vec{k},\vec{k}'\in\text{BZ}}
	d^\dagger_{\text{BZ}(\vec{k}+\vec{q}),\alpha} d^\dagger_{\vec{k}',\gamma} d^\pdagger_{\vec{k},\beta} d^\pdagger_{\text{BZ}(\vec{k}'+\vec{q}),\delta} \mathcal{F}_{\alpha\beta}(\vec{k},\vec{q})  \mathcal{F}_{\gamma\delta}\left(\text{BZ}(\vec{k}'+\vec{q}),-\vec{q}\right)\nonumber \\
	&= \quad\sum_{\vec{q}\notin\text{RL}} V(\vec{q}) \sum_{\vec{k},\vec{k}'\in\text{BZ}}
	d^\dagger_{\text{BZ}(\vec{k}+\vec{q}),\alpha} d^\dagger_{\vec{k}',\gamma} d^\pdagger_{\text{BZ}(\vec{k}'+\vec{q}),\delta} d^\pdagger_{\vec{k},\beta} \mathcal{F}_{\alpha,\beta}(\vec{k},\vec{q})    \mathcal{F}_{\gamma\delta}^\dagger(\vec{k}',\vec{q}).
\end{align}
After mean-field decoupling,
\begin{equation}
\begin{aligned}
	&\hat{H}_1^{\text{eff}} = \sum_{\vec{q}\notin\text{RL}} V(\vec{q}) \sum_{\vec{k},\vec{k}'\in\text{BZ}}
	\mathcal{F}_{\alpha\beta}(\vec{k},\vec{q}) \mathcal{F}_{\gamma\delta} ^\dagger(\vec{k}',\vec{q})\times \\ & \times
\left(\braket{d^\dagger_{\vec{k}',\gamma}
		d^\pdagger_{\text{BZ}(\vec{k}'+\vec{q}),\delta}}
	d^\dagger_{\text{BZ}(\vec{k}+\vec{q}),\alpha}  d^\pdagger_{\vec{k},\beta}
	\delta^\pdagger_{\vec{k}', \text{BZ}(\vec{k}'+\vec{q})} 
	+\braket{d^\dagger_{\text{BZ}(\vec{k}+\vec{q}),\alpha} 
		d^\pdagger_{\vec{k},\beta}}
	d^\dagger_{\vec{k}',\gamma} 
	d^\pdagger_{\text{BZ}(\vec{k}'+\vec{q}),\delta} 
	\delta^\pdagger_{\text{BZ}(\vec{k}+\vec{q}), \vec{k}}+ \right. \\
	&\left.
	- \braket{d^\dagger_{\text{BZ}(\vec{k}+\vec{q}),\alpha} d^\pdagger_{\text{BZ}(\vec{k}'+\vec{q}),\delta}} d^\dagger_{\vec{k}',\gamma} d^\pdagger_{\vec{k},\beta}
	\delta^\pdagger_{\text{BZ}(\vec{k}+\vec{q}), \text{BZ}(\vec{k}'+\vec{q})} - \braket{d^\dagger_{\vec{k}',\gamma} d^\pdagger_{\vec{k},\beta}}
	d^\dagger_{\text{BZ}(\vec{k}+\vec{q}),\alpha}
	d^\pdagger_{\text{BZ}(\vec{k}'+\vec{q}),\delta} 
	\delta^\pdagger_{\vec{k}', \vec{k}}
	\right).
	\label{NormalOrderedMeanFieldExpression}
\end{aligned}
\end{equation}
The Hartree contribution is zero for the fermionic model, since $\delta_{\text{BZ}(\vec{k}+\vec{q}),\vec{k}} = 0 ,\forall \vec{k}\in \text{BZ}\; \text{and} \; \vec{q} \notin \text{RL}$. Therefore, with the Fock contribution as the only remaining part, the effective Hamiltonian takes the form
\begin{align}
	\hat{H}_1^{\text{eff}} = -&\sum_{\vec{q}\in\mathbbm{R}} V(\vec{q}) \sum_{\vec{k}\in\text{BZ}}
	\mathcal{F}_{\alpha\beta}(\vec{k},\vec{q}) \mathcal{F}_{\gamma\delta}^\dagger(\vec{k},\vec{q}) \nonumber \big[
	(P_{\text{BZ}(\vec{k}+\vec{q})})^\pdagger_{\alpha, \delta} d^\dagger_{\vec{k},\gamma} d^\pdagger_{\vec{k},\beta} + (P_{\vec{k}})^\pdagger_{\gamma, \beta}
	d^\dagger_{\text{BZ}(\vec{k}+\vec{q}),\alpha}
	d^\pdagger_{\text{BZ}(\vec{k}+\vec{q}),\delta} 
	\big]
	\label{finalMFHamiltonianComponentWise}
\end{align}
with the projector defined as $\left(P_{\text{BZ}\left(\vec{k}+\vec{q}\right)}\right)_{\alpha,\delta}=\braket{d^\dagger_{\text{BZ}(\vec{k}+\vec{q}),\alpha} d^\pdagger_{\text{BZ}(\vec{k}'+\vec{q}),\delta}}$. We also used the fact that $\delta_{\vec{k}', \vec{k}}=\delta^\pdagger_{\text{BZ}(\vec{k}+\vec{q}), \text{BZ}(\vec{k}'+\vec{q})}$.
Switching back to matrix notation in
\begin{align}
	\hat{H}_1^{\text{eff}} = -&\sum_{\vec{q}\in\mathbbm{R}} V(\vec{q}) \sum_{\vec{k}\in\text{BZ}}
	\big[
	d^\dagger_{\vec{k}} \mathcal{F}^\dagger(\vec{k},\vec{q}) (P_{\text{BZ}(\vec{k}+\vec{q})})^T \mathcal{F}(\vec{k},\vec{q})^\pdagger d^\pdagger_{\vec{k}} + 
	d^\dagger_{\text{BZ}(\vec{k}+\vec{q})}
	\mathcal{F}(\vec{k},\vec{q})
	(P_{\vec{k}})^T
	\mathcal{F}^\dagger(\vec{k},\vec{q})
	d^\pdagger_{\text{BZ}(\vec{k}+\vec{q})} 
	\big],
\end{align}
and taking another index shift in the second term $\vec{q} \rightarrow -\vec{q}$ and $\vec{k} \rightarrow \text{BZ}(\vec{k}+\vec{q})$ yields
\begin{equation}
	\hat{H}_1^{\text{eff}} = -2 \sum_{\vec{q}\in\mathbbm{R}} V(\vec{q}) \sum_{\vec{k}\in\text{BZ}}
	d^\dagger_{\vec{k}}
	\left[ \mathcal{F}^\dagger(\vec{k},\vec{q}) (P_{\text{BZ}(\vec{k}+\vec{q})})^T \mathcal{F}(
    (\vec{k},\vec{q}) \right]
	d^\pdagger_{\vec{k}}
	.
\end{equation}

\begin{figure}[t!]
    \centering
    \includegraphics[width=1\textwidth]{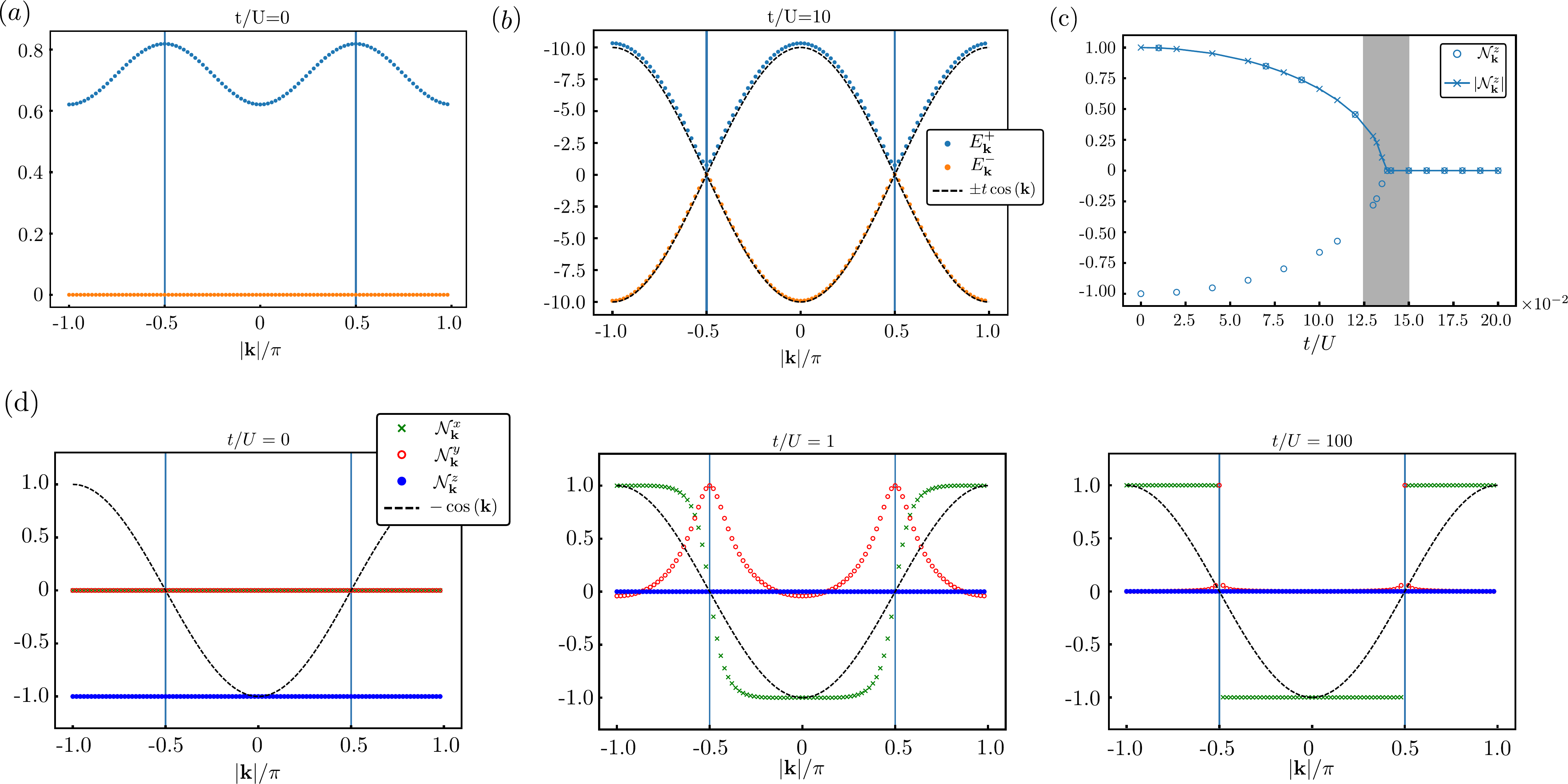}
    \caption{Observables for the fermionic model from the perspective of Hartree-Fock for $N_{e}=100$ electrons. Band structure for the fermionic model at $t/U=0$ (a) and  $t/U=10$ (b) obtained according to the expression \eqref{eq:hfunit}. (c) Order parameter $\xi$ as a function of $t/U$. (d) Momentum-resolved fermionic bilinears (according to \equref{eq:observablen}) as a function of $\vec{k}$ for different values of $t/U$.}
    \label{fig:sihf1}
\end{figure}

To summarize, within mean-field theory, we have found an effective Hamiltonian to the interacting term given by the Hatree-Fock equations 
\begin{equation}
	\begin{split}
		\hat{H}^{\text{eff}} &=
		\sum_{\vec{k}\in\text{BZ}}
		d^\dagger_{\vec{k}} (f[P_{\vec{k}}] + h_{\vec{k}}) d^\pdagger_{\vec{k}} \\
		f[P_{\vec{k}}] &= -2 \sum_{\vec{q}\notin\text{RL}} V(\vec{q}) 
		\big( \mathcal{F}^\dagger(\vec{k},\vec{q}) (P_{\text{BZ}(\vec{k}+\vec{q})})^T 	\mathcal{F}(\vec{k},\vec{q})^\pdagger \big) \\
		h_{\vec{k}} &= \sum_{\vec{q}\notin\text{RL}}
		V(\vec{q}) \mathcal{F}^\dagger(\vec{k},\vec{q}) \mathcal{F}(\vec{k},\vec{q})^\pdagger + t \cos(\vec{k}) \sigma^\pdagger_x.
	\end{split}
	\label{eq:hmfDefiniton}
\end{equation}
which needs to be solved self-consistently. The HF ground state energy is then  given by
\begin{equation}
	E_{\text{HF}} = \sum_{\vec{k} \in \text{BZ}} \left(\left(f[P_{\vec{k}}] + h_{\vec{k}}\right)_{-, -} - E_{\vec{k}}^{0}\right)
\end{equation}
with the energy offset
\begin{equation}
	\begin{split}
	E^{0}_{\vec{k}} &= \frac{1}{2} \tr{\left[P_{\vec{k}}^T\right]} f[P_{\vec{k}}]= -\sum_{\vec{q}\notin\text{RL}} V(\vec{q}) \tr{\left[P_{\vec{k}}^T \mathcal{F}^\dagger(\vec{k},\vec{q})(P_{\text{BZ}(\vec{k}+\vec{q})})^T \mathcal{F}(\vec{k},\vec{q})\right]}.
	\end{split}
\end{equation}

To solve equations \equref{eq:hmfDefiniton}, we first consider an initial ansatz for $P$. This can be chosen to correspond to the (i) the interacting limit $t/U\rightarrow 0$ with $P_{\vec{k}}^{ \text{chiral}}=\text{diag}\left(0,1\right)$, (ii) the kinetic edge case $P_{\vec{k}}^{\text{band}}=\left(\sigma_{0}-\sigma_{x}\right)/2$ or (iii) to a random matrix $P_{\vec{k}}^{\text{rand}}$ that fulfills the projector properties $P_{\vec{k}} = P_{\vec{k}}^\dagger = (P_{\vec{k}})^2$. In all simulations for this work we always choose (iii). 

The Hermitian matrix $(f[P_{\vec{k}}] + h_{\vec{k}})$ in \equref{eq:hmfDefiniton} is diagonalized in each iteration step according to 
\begin{equation}
	D_{\vec{k}} = U_{\vec{k}}^{-1} (f[P_{\vec{k}}] + h_{\vec{k}}) U^\pdagger_{\vec{k}}, \quad 	D_{\vec{k}} =
	\begin{pmatrix}
		E_{\vec{k}}^+ & 0 \\
		0 & E_{\vec{k}}^- 
	\end{pmatrix}\quad \text{and} \quad 	U_k =
	\begin{pmatrix}
		v_0^+ & v_0^- \\
		v_1^+ & v_1^- 
	\end{pmatrix},
	\label{eq:hfunit}
\end{equation}
where we ensure that $E_{\vec{k}}^{-} \le E_{\vec{k}}^{+}$ in the $\vec{k}$-dependent diagonal matrix $D_{\vec{k}}$, and the unitary matrix $U_{\vec{k}}$ is composed by the normalized eigenvalues $v^+$ and $v^-$. We initialize $D_{\vec{k}}=\text{diag}\left(0,1\right)$, which corresponds to filling the lower energy band of the mean-field Hamiltonian $\hat{H}_{\text{HF}}$ and calculate the projector of the next iteration step $P_{\vec{k}}^{i+1} = U_{\vec{k}}^{*}D U_{\vec{k}}^T$ which, in turn, defines $f[P_{\vec{k}}^{i+1}]$. This is done until the projector converges according to $\|P^{(i+1)}_{\vec{k}}- P_{\vec{k}}^i\|_F < 10^{-10}$ where $\|\cdot\|_F$ denotes the Frobenius norm.

After convergence we can also evaluate other observables besides the ground state energy, like the momentum-resolved fermionic bilinears according to \equref{eq:observablen}. Within HF, in 
\figref{fig:sihf1} we show the band structure (a,b,c) ($\epsilon_{\vec{k}, p}$ in  \equref{eq:unitarytransf}) and the fermionic bilinears in the chiral basis (d, e), which illustrate the metal-insulator phase transition described in the main text. For small $t/U$ we see a clear band gap in the insulating regime (\figref{fig:sihf1}a). The energy gap gets smaller for intermediate $t/U$, and vanishes for large $t/U$ (c) with the appearance of two Dirac cones around $|\vec{k}|=\pm \pi/2$. By defining the order parameter $\xi=\sum_{\vec{k}}\mathcal{N}_{\vec{k}}^{z}/N_{e}$, we see that the metal-insulator phase transition is set around $t=0.14$ (d). In \figref{fig:sihf1}e, we also note that  at $t/U=0$,  $\braket{\mathcal{N}_{\vec{k}}^{x}}=0$ while large $t/U$ leads to $|\braket{\mathcal{N}_{\vec{k}}^{x}}|=1$ with sign changes at $|\vec{k}|=\pm\pi/2$.

\section{Variational Monte Carlo}
\label{sec:vmc}
\subsection{Calculating Observables}
According to the ansatz \equref{eq:tqsmod}, the expectation value 
\begin{equation}
    \begin{aligned}
       &\frac{\langle\Psi_{\{\vec{\theta},\alpha\}}|O| \Psi_{\{\vec{\theta},\alpha\}}\rangle}{\langle\Psi_{\{\vec{\theta},\alpha\}} \mid \Psi_{\{\vec{\theta},\alpha\}}\rangle} =\frac{1}{\alpha^{2}+\left(1-\alpha^{2}\right)\sum_{\vec{s}^{\prime \prime}\neq \text{RS}}|\psi_{\vec{\theta}}(\vec{s}^{\prime \prime})|^2}\left[\alpha^{2}\bra{\text{RS}}\hat{O}\ket{\text{RS}}+\alpha\sqrt{1-\alpha^{2}}\sum_{\vec{s}\neq{\text{RS}}} \braket{\text{RS}|\hat{O}|\vec{s}} \psi_{\vec{\theta}}\left(\vec{s}\right) + \right. \\ &  \left. + \sum_{\vec{s}\neq\text{RS}}\left(\left(1-\alpha^{2}\right)\sum_{\vec{s}^{\prime}\neq \text{RS}} \braket{\vec{s}|\hat{O}|\vec{s}^{\prime}} \psi_{\vec{\theta}}\left(\vec{s}^{\prime}\right)\psi^{*}_{\vec{\theta}}\left(\vec{s}\right)
      +
      \alpha\sqrt{1-\alpha^{2}}\braket{\vec{s}|\hat{O}|\text{RS}} \psi_{\vec{\theta}}^{*}\left(\vec{s}\right)\right) \right]\\ 
      &=\frac{1}{\mathbb{N}_{\{\vec{\theta},\alpha\}}}\left[\alpha^{2}O_{\text{RS}} + \left(1-\alpha^{2}\right)\sum_{\vec{s},\vec{s}^{\prime}\neq \text{RS}} \left|\psi_{\vec{\theta}}\left(\vec{s}\right)\right|^{2} \braket{\vec{s}|\hat{O}|\vec{s}^{\prime}}\frac{\psi_{\vec{\theta}}\left(\vec{s}^{\prime}\right)}{\psi_{\vec{\theta}}\left(\vec{s}\right)}
      +
      \alpha\sqrt{1-\alpha^{2}}\sum_{\vec{s}\neq{\text{RS}}}\left|\Psi_{\vec{\theta}}(\vec{s})\right|^2 \left( O_{\vec{s} \text{RS}} \frac{1}{\Psi_{\vec{\theta}}\left(\vec{s}\right)} +O_{\text{RS}\vec{s}}\frac{1}{\Psi^{*}
      _{\vec{\theta}}\left(\vec{s}\right)} \right) \right] \\
           &=\frac{1}{\mathbb{N}_{\{\vec{\theta},\alpha\}}}\left[\alpha^{2}O_{\text{RS}} + \left(1-\alpha^{2}\right)\sum_{\vec{s}\neq \text{RS}} \left|\Psi_{\vec{\theta}}\left(\vec{s}\right)\right|^{2} O_{\text{loc}}^{\vec{s}^{\prime}\neq\text{RS}}\left(\vec{s}\right) 
      +
      \alpha\sqrt{1-\alpha^{2}}2\Re\left(\sum_{\vec{s}\neq{\text{RS}}}\left|\Psi_{\vec{\theta}}(\vec{s})\right|^2  O_{\vec{s} \text{RS}} \frac{1}{\Psi_{\vec{\theta}}\left(\vec{s}\right)}\right)\right],
    \end{aligned}
\end{equation}
which is equivalent to \equref{eq:expenergy} for the expectation value of $\hat{H}$. 
Note that the normalization factor $\mathbb{N}_{\{\vec{\theta},\alpha\}}=\alpha^{2}+\left(1-\alpha^{2}\right)\sum_{\vec{s}^{\prime \prime}\neq \text{HF}}|\Psi_{\vec{\theta}}(\vec{s}^{\prime \prime})|^2\rightarrow 1$ for any $\alpha$ if the $\text{RS}$ is not sampled from the TQS.
Likewise, for the gradient of the energy functional
\begin{equation}
\begin{aligned}
   \nabla_{\vec{\theta}}E\left(\vec{\theta}, \alpha \right) &= \partial_{\vec{\theta}}\left[\sum_{\vec{s},\vec{s}^{\prime}\neq\text{HF}}\left(1-\alpha^{2}\right)\Psi_{\vec{\theta}}^{*}\left(\vec{s}\right)H_{\vec{s}\vec{s}^{\prime}}\Psi_{\vec{\theta}}\left(\vec{s}^{\prime}\right)+\alpha\sqrt{1-\alpha^{2}}\sum_{\vec{s}\neq\text{HF}}\left(\Psi_{\vec{\theta}}^{*}\left(\vec{s}\right)H_{\vec{s}\text{HF}}+H_{\text{HF}\vec{s}}\Psi_{\vec{\theta}}\left(\vec{s}\right)\right)\right] \\ 
    &= \left(1-\alpha^{2}\right)\sum_{\vec{s},\vec{s}^{\prime}\neq\text{HF}}\left(\partial_{\vec{\theta}}\Psi^{*}_{\vec{\theta}}\left(\vec{s}\right)\Psi_{\vec{\theta}}\left(\vec{s}^{\prime}\right)\frac{\Psi^{*}_{\vec{\theta}}\left(\vec{s}\right)}{\Psi^{*}_{\vec{\theta}}\left(\vec{s}\right)}\frac{\Psi_{\vec{\theta}}\left(\vec{s}\right)}{\Psi_{\vec{\theta}}\left(\vec{s}\right)}+
    \Psi^{*}_{\vec{\theta}}\left(\vec{s}\right)\partial_{\vec{\theta}}\Psi_{\vec{\theta}}\left(\vec{s}^{\prime}\right)\frac{\Psi^{*}_{\vec{\theta}}\left(\vec{s}^{\prime}\right)}{\Psi^{*}_{\vec{\theta}}\left(\vec{s}^{\prime}\right)}\frac{\Psi_{\vec{\theta}}\left(\vec{s}^{\prime}\right)}{\Psi_{\vec{\theta}}\left(\vec{s}^{\prime}\right)}\right)H_{\vec{s}\vec{s}^{\prime}} +\\ 
    &  +\alpha\sqrt{1-\alpha^{2}}\sum_{\vec{s}\neq\text{HF}}\left(\partial_{\vec{\theta}}\Psi_{\vec{\theta}}^{*}\left(\vec{s}\right)H_{\vec{s}\text{HF}}\frac{\Psi_{\vec{\theta}}\left(\vec{s}\right)}{\Psi_{\vec{\theta}}\left(\vec{s}\right)}\frac{\Psi^{*}_{\vec{\theta}}\left(\vec{s}\right)}{\Psi^{*}_{\vec{\theta}}\left(\vec{s}\right)}+H_{\text{HF}\vec{s}}\partial_{\vec{\theta}}\Psi_{\vec{\theta}}\left(\vec{s}\right)\frac{\Psi_{\vec{\theta}}\left(\vec{s}\right)}{\Psi_{\vec{\theta}}\left(\vec{s}\right)}\right)\\ 
        &= \left(1-\alpha^{2}\right)2\Re
 \left \langle \partial_{\vec{\theta}}\log \Psi^{*}_{\vec{\theta}}\left(\vec{s}\right) H_{\text{loc}}^{\vec{s}^{\prime}\neq\text{HF}}\left(\vec{s}\right) \right\rangle + \\ 
 &+
      \alpha\sqrt{1-\alpha^{2}}\left(\left\langle \partial_{\vec{\theta}}\log \Psi^{*}_{\vec{\theta}}\left(\vec{s}\right)H_{\vec{s}\text{HF}}\frac{1}{\Psi_{\vec{\theta}}\left(\vec{s}\right) } \right\rangle + \sum_{\vec{s}\neq\text{HF}}\partial_{\vec{\theta}}\log \Psi_{\vec{\theta}}\left(\vec{s}\right)H_{\text{HF}\vec{s}}\Psi_{\vec{\theta}}\left(\vec{s}\right) \right),\\
         &= \left(1-\alpha^{2}\right)2\Re
 \left \langle \partial_{\vec{\theta}}\log \Psi^{*}_{\vec{\theta}}\left(\vec{s}\right) H_{\text{loc}}^{\vec{s}^{\prime}\neq\text{HF}}\left(\vec{s}\right) \right\rangle+
      \alpha\sqrt{1-\alpha^{2}}2\Re\left\langle \partial_{\vec{\theta}}\log \Psi^{*}_{\vec{\theta}}\left(\vec{s}\right)H_{\vec{s}\text{HF}}\frac{1}{\Psi_{\vec{\theta}}\left(\vec{s}\right) } \right\rangle,  \\  
         &= 2\Re\left[
 \left \langle \partial_{\vec{\theta}}\log \Psi^{*}_{\vec{\theta}}\left(\vec{s}\right)\left( \left(1-\alpha^{2}\right) H_{\text{loc}}^{\vec{s}^{\prime}\neq\text{HF}}\left(\vec{s}\right)+
      \alpha\sqrt{1-\alpha^{2}}H_{\vec{s}\text{HF}}\frac{1}{\Psi_{\vec{\theta}}\left(\vec{s}\right) }\right) \right\rangle\right],    
\end{aligned}
\end{equation}
in accordance with  \equref{eq:gradtheta} in the main text.

\subsection{Performance Analysis of Optimizers and Hyperparameters}

\begin{figure}[h]
    \centering
    \includegraphics[width=0.5\textwidth]{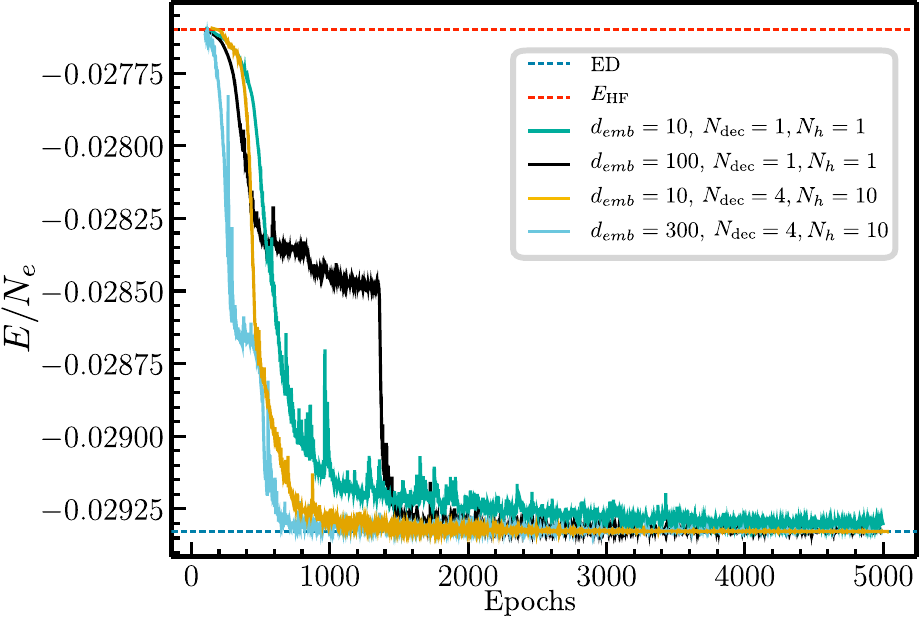}
    \caption{Ground state energy per site as a function of epochs for distinct combinations of $d_{\text{emb}}$, $N_{\text{dec}}$ and $N_{h}$ at $t/U=0.10$ for $N_{e}=10$. The dashed blue (red) line indicates the ground state energy obtained from ED (HF).}
    \label{fig:comparisonembeddim}
\end{figure}

The main hyperparameters that define a decoder-only Transformer architecture are given by $N_{\text{dec}}$, $d_{\text{emb}}$ and, $N_{\text{h}}$ which stand, respectively, for the number of encoding layers, embedding dimension and number of attention heads (see \figref{fig:generalmethodology}b). For a more in depth description of this architecture, we refer the reader to the references \cite{Vaswani2017Jun, Lin2021Jun}. More importantly, the representational power of these architectures is directly related to these quantities \cite{sanfordRepresentationalStrengthsLimitations2023}. In \figref{fig:comparisonembeddim} we show how the training of the Transformers with the HF-basis is affected in terms of different combinations of these parameters. We focus on $N_{e}=10$ electrons, but analogous results can  be directly obtained for bigger system sizes. For $t/U=0.10$, we see that the training can be made faster and more accurate by solely solely increasing $d_{\text{emb}}$, or if kept fixed, by increasing $N_{\text{dec}}$ and $N_{h}$. Similarly, we observed the same behavior for $t/U \in [0.01,0.04] \cup [0.10,0.20]$. Therefore, we used the combination $d_{\text{emb}}=300$, $N_{\text{dec}}= 4$ and $N_{h}=10$ for the plots in \figref{fig:scaling}a and \figref{fig:bases}. Additionally, For \figref{fig:scaling}, the results were obtained for  $n_{\text{epochs}}=2\times 10^{4}$.

For the optimization of the $\vec{\theta}$ parameters, the learning rate $\lambda_{\vec{\theta}}$ is changed according to the following scheduler \cite{zhang2023tqs} 
\begin{equation}
\lambda_{\vec{\theta}}^{i+1} = \lambda_{\vec{\theta}}^{i}+\beta d_{\text{emb}}^{-0.5} \min \left(i_{\text {step }}^{-0.75}, i_{\text {step }} i_{\text {warmup}}^{-1.75}\right)
\end{equation}
with $\beta=0.2$ representing a scale factor, $i_{\text{step}}$ the current iteration index, and $i_{\text{warmup}}=700$
the number of warming up steps.  This scheduler allows the learning rate to be increased linearly during the first $i_{warmup}$ epochs, and then decreased polynomially during the remaining $\left(n_{\text{epochs}}-i_{warmup}\right)$ steps \cite{vaswaniAttentionAllYou2017}. Although a similar learning rate scheduler can be imposed for the optimization of $\alpha$, according to \equref{eq:gradalpha}, we noticed that the Transformer converges better for $t/U<0.10$ with a constant learning rate. The best converged energies between $\lambda_{\alpha_{0}}=100$ (for intermediate $t/U$) and $\lambda_{\alpha_{0}}=2$ (for small $t/U$) are shown for the simulations in \figref{fig:scaling} and \figref{fig:bases}.

\begin{figure}[t]
    \centering
    \includegraphics[width=1\textwidth]{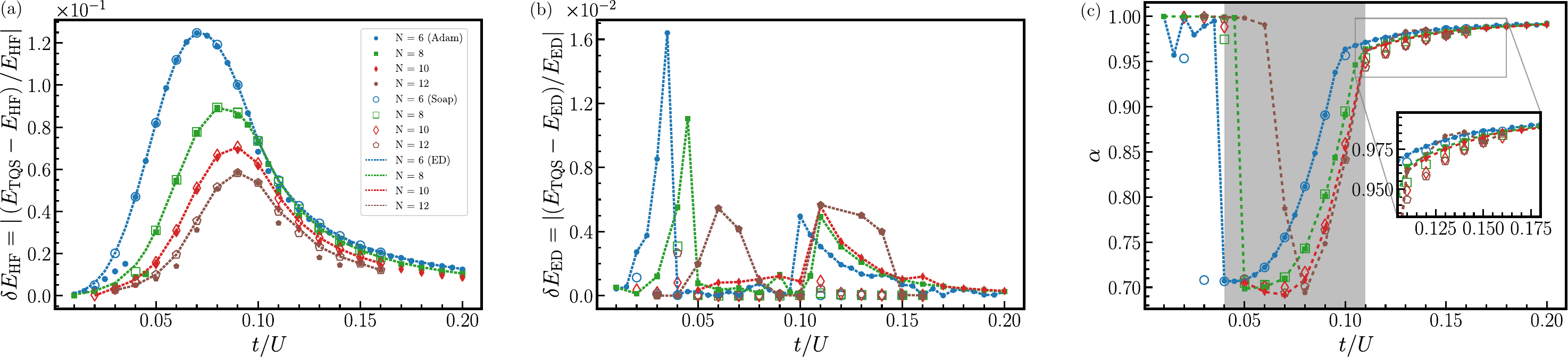}
    \caption{Comparison of the performance of the HF-TQS when using ADAM \cite{Kingma2014Dec} and SOAP \cite{Vyas2024Sep} as a function of $t/U$ for different system sizes. (a) Relative difference between the converged ground state energy per $N_{e}$ and HF for $x$ as ED (solid lines) and TQS (markers) in $\delta E_{\text{HF}}=\left|\left(E_{x}-E_{\text{HF}}\right)/E_{\text{HF}}\right|$ with SOAP (unfilled markers) and ADAM (filled markers). (b) The absolute value of the relative error $\delta E_{\text{ED}}=\left|(E_{\text{TQS}}-E_{\text{ED}})/E_{\text{ED}}\right|$. The corresponding converged $\alpha$ weights (according to \equref{eq:expenergy}) are shown in (b). The gray region indicates the surrounding region to the metal-insulator transition. }
    \label{fig:comparisonoptimizers}
\end{figure}

Most importantly, we noticed that the choice of the specific stochastic optimizer was the most important point for a smooth and consistent convergence for different values of $t/U$ for the fermionic model. More specifically, 
adaptative moment estimation (ADAM) \cite{Kingma2014Dec} and other variations of the simple stochastic gradient descent method tended to converge to ground state energies with less accuracy away from the critical region (see \figref{fig:comparisonoptimizers}), irrespective of different combinations of the previously mentioned hyperparameters. We noticed a significant improvement when considering preconditioned gradient methods \cite{Gupta2018Feb,Vyas2024Sep} in these regions. Likewise, there are also deviations for the converged $\alpha$ using these different optimizers, specially around the borders of the gray region in \figref{fig:comparisonoptimizers}c.

\begin{figure}[h]
    \centering
    \includegraphics[width=1\textwidth]{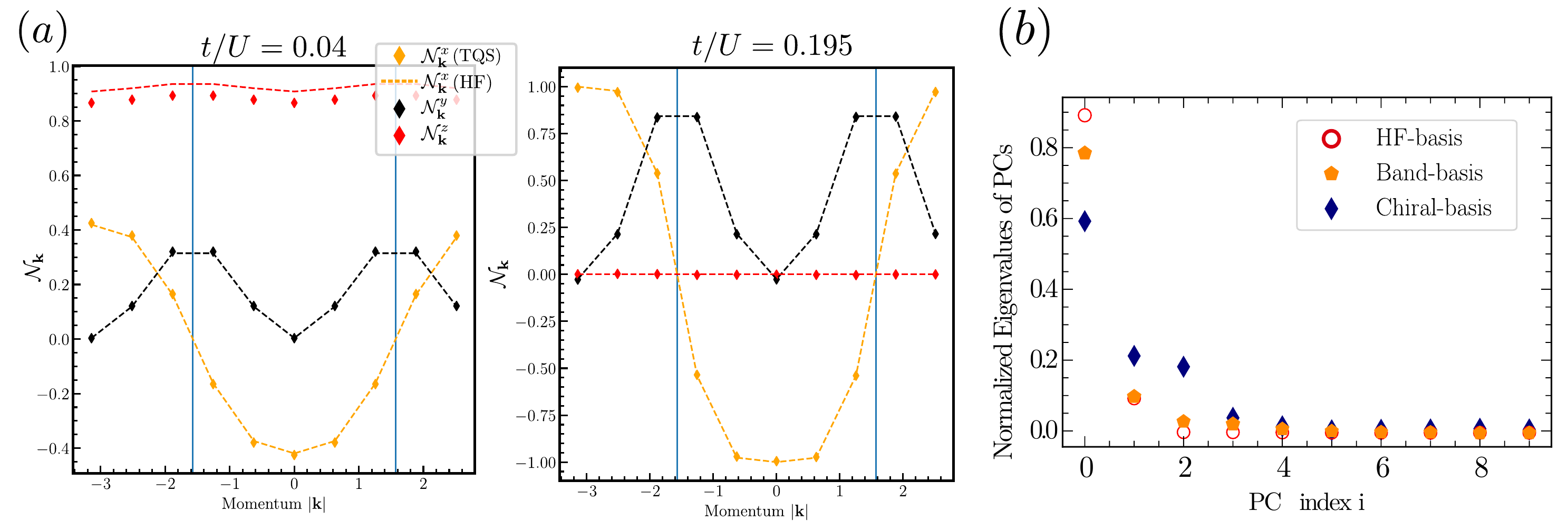}
    \caption{HF-TQS results  (markers) for the momentum-resolved fermionic bilinears $\mathcal{N}^j_{\vec{k}}$, defined in \equref{eq:observablen},  obtained from HF (dashed lines) for $N_{e}=10$ at $t/U=0.04$ (left) and $t/U=0.195$ (right). (b) Normalized eigenvalues of the first ten principal components from the PCA analysis shown in  \figref{fig:pcalatent} ($t/U=0.12$). Results for the three distinct bases (indicated by markers) defined in the main text.}
    \label{fig:bilinearsvmc}
\end{figure}

\subsection{Further results}
In this section we present additional results complementing the discussions in \secref{sec:results}. We focus on $N_{e}=10$ electrons, although similar results can be obtained for bigger system sizes.

Figure \ref{fig:bilinearsvmc}a demonstrates that corrections to HF for the expectation values of the fermionic bilinears, c.f. \equref{eq:observablen}, diminish away from the critical region, in accordance with \secref{subsec:obs}.

Our PCA analysis of the latent space, shown in \figref{fig:pcalatent} of the main text, reveals that the first two principal components account for at least $ 80\%$ of the total data variation across all bases (see \figref{fig:bilinearsvmc}b). This high percentage of explained variance remained consistent, with slight variations, for other coupling parameters such as  $t/U=0.04$ and $t/U=0.16$ in \figref{fig:pcalatent}.  

\begin{figure}[htp!]
    \centering
    \includegraphics[width=1\textwidth]{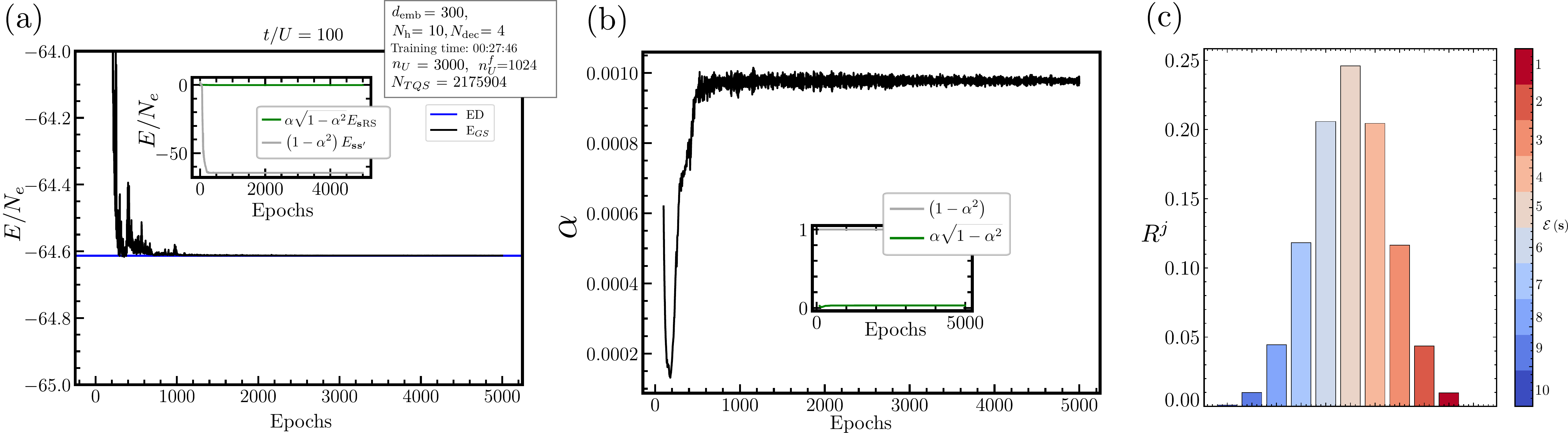}
    \caption{Training plots for the Transformer with the band-basis for $N_{e}=10$ and $t/U=100$.  Ground state per site (a) and reweighting factor $\alpha$ (b) as a function of epochs. The inset in (a) shows the different contributions to the ground state energy according to \equref{eq:expenergy}. (c) Histogram with the total relative frequency, according to eq. \eqref{eq:totalweight}, for the excitation classes $\mathcal{E}\left(\vec{s}\right)$ from \equref{eq:labeldof}.}
    \label{fig:bigtbandbasis}
\end{figure}

Finally, we also examined the Transformer's convergence in the band basis for large $t/U$. As shown in \figref{fig:bigtbandbasis}(a,b), even though the fully polarized RS is not a good representative of the ground state in this regime (see \figref{fig:sihf1}c), the Transformer is still able to achieve good agreement with ED (\figref{fig:sihf1}a), as stated in \secref{subsec:otherbases}. 

\end{appendix}

\end{document}